\documentclass[final,authoryear,3p,11pt,times]{elsarticle}

\usepackage{afterpage}
\usepackage{algorithm}
\usepackage{algpseudocode}
\usepackage{amsmath}
\usepackage{amssymb}
\usepackage{amsthm}
\usepackage{bm}
\usepackage{booktabs}
\usepackage{calc}
\usepackage{caption}
\usepackage{subcaption}
\usepackage[table]{xcolor} 
\usepackage{amsmath,amssymb}
\usepackage{booktabs}
\usepackage{tabularx}

\usepackage{enumerate}
\usepackage{graphicx}
\usepackage{ifthen}
\usepackage[parfill]{parskip} 
\usepackage{multicol}
\usepackage{siunitx}
\usepackage{ulem}

\usepackage[colorlinks]{hyperref}
\usepackage[nameinlink,capitalise]{cleveref}
\crefname{appendix}{}{}

\usepackage{tikz}
\usetikzlibrary{arrows}
\usetikzlibrary{automata}
\usetikzlibrary{calc}
\usetikzlibrary{chains}
\usetikzlibrary{positioning}
\usetikzlibrary{shapes}
\usepackage{setspace} 
\usepackage{varwidth}

\usepackage{tcolorbox}
\newtcolorbox[auto counter]{problem}[2][]{colframe=blue!30, colback=blue!5, coltitle=black, title=Problem~\thetcbcounter~ #2,#1}

\usepackage{spverbatim} 
\usepackage{listings} 

\definecolor{Demiray}{rgb}{0.1725, 0.6275, 0.1725}
\definecolor{Gent}{rgb}{1.0, 0.4902, 0.0431}
\definecolor{Holzapfel}{rgb}{0.1216, 0.4667, 0.7059}
\definecolor{MooneyRivlin}{rgb}{0.8392, 0.1529, 0.1569}
\definecolor{Ogden}{rgb}{0.5804, 0.4039, 0.7412}
\definecolor{BlatzKo}{rgb}{0.5490, 0.3373, 0.2941}
\definecolor{NeoHooke}{rgb}{0.7373, 0.7412, 0.1333}

\usepackage{amsmath,amssymb,amsthm}  
\usepackage{bm}                      
\usepackage{mathtools}               
\usepackage{enumitem}                
\usepackage{cleveref}                
\usepackage{geometry}
\edef\svtheparindent{\the\parindent}
\usepackage{parskip}
\parindent=\svtheparindent\relax

\newcommand{\MF}{Material Fingerprinting}

\definecolor{ForestGreen}{RGB}{34,139,34}
\definecolor{InternationalOrange}{rgb}{1.0, 0.31, 0.0}
\definecolor{WineRed}{RGB}{139,0,0}

\newcommand{\CHANGE}[1]{\textcolor{black}{#1}} 
\newcommand{\CHANGETWO}[1]{\textcolor{black}{#1}} 

\newcommand{\boldface}[1]{\boldsymbol{#1}}  

\newcommand{\bff}{\boldface{f}}

\newcommand{\bfu}{\boldface{u}}

\newcommand{\bfC}{\boldface{C}}

\newcommand{\bfF}{\boldface{F}}

\newcommand{\bfI}{\boldface{I}}

\newcommand{\bfQ}{\boldface{Q}}

\newcommand{\bfalpha}{\boldsymbol{\alpha}}

\newcommand{\bftheta}{\boldsymbol{\theta}}

\newcommand{\Rset}{\mathbb{R}}

\newcommand{\be}{\begin{equation}}
\newcommand{\ee}{\end{equation}}
\newcommand{\bea}{\begin{equation}\begin{aligned}}
\newcommand{\eea}{\end{aligned}\end{equation}}
\newcommand{\beq}{\begin{eqnarray}}
\newcommand{\eeq}{\end{eqnarray}}
\newcommand{\bem}{\begin{multline}}
\newcommand{\eem}{\end{multline}}
\newcommand{\ba}{\begin{align}}
\newcommand{\ea}{\end{align}}
\newcommand{\bcase}{\left\{ \begin{array}{ll}}
\newcommand{\ecase}{\end{array} \right.}
\begin{document}

\begin{frontmatter}

\title{Unsupervised Material Fingerprinting: Ultra-fast hyperelastic model discovery from full-field experimental measurements}

\author[fau]{Moritz Flaschel\corref{cor1}\fnref{contrib}}
\ead{moritz.flaschel@fau.de}
\author[fau]{Miguel Angel Moreno-Mateos\corref{cor1}\fnref{contrib}}
\ead{miguel.moreno@fau.de}
\author[fau]{Simon Wiesheier\fnref{contrib}}
\author[fau,gal]{Paul Steinmann}
\author[fau,stan]{Ellen Kuhl}

\fntext[contrib]{M. Flaschel, M. A. Moreno-Mateos, and S. Wiesheier contributed equally to this work.}
\cortext[cor1]{Corresponding authors}

\address[fau]{Institute of Applied Mechanics, Friedrich-Alexander-Universität Erlangen-Nürnberg, Egerlandstr. 5, 91058, Erlangen, Germany}
\address[stan]{Department of Mechanical Engineering, 318 Campus Drive, Stanford University, California 94305, United States.}
\address[gal]{Glasgow Computational Engineering Centre, School of Engineering, University of Glasgow, G12 8QQ, United Kingdom}

\begin{abstract}
Institutions such as DIN, EN, ISO, and ASTM define standardized experimental protocols to ensure the reproducible and consistent characterization of mechanical material behavior.
If a comprehensive database of precomputed material responses were available for these standardized tests, material characterization could instead be formulated as a pattern recognition problem, which enables significantly faster and more robust identification compared to optimization-based methods.
This paradigm underlies our recently proposed \MF{} method.
\MF{} is a lookup table-based strategy to \CHANGE{infer} material models from experimental measurements, which completely avoids the need to solve a \CHANGE{continuous optimization problem}.
In an offline phase, a comprehensive database of simulated material responses, so-called material fingerprints, is generated for a predefined, standardized experimental setup. Although it can be extended a posteriori, the database is generated only once. Then, it is used repeatedly in an ultra-fast online phase to \CHANGE{infer} material models from experiments performed according to guidelines and standardized sample geometries.
The experimentally measured fingerprint is compared with a database to identify the closest match.
The main advantages of \MF{} are two: i) ultra-fast material model \CHANGE{inference} in just a few seconds and ii) robustness in the identification because a continuous optimization problem does not need to be solved. The method circumvents ill-posedness and non-convex landscape-related issues in traditional, usually computationally-expensive, methods.
To date, there exists no demonstration of \MF{} applied to unsupervised experimental datasets (i.e., sets of full-field displacements and global reaction forces). Here, unsupervised \MF{} offers a robust material modeling framework by directly comparing precomputed simulated displacements and reaction forces with experimental counterparts.
In this work, we apply this strategy to biaxial deformation tests of soft elastomer specimens (Elastosil, Sylgard, and VHB tape) with a central strain concentrator and inhomogeneity in the deformation field. We construct a single database across different materials and \CHANGE{infer} hyperelastic material models.
We show that, for an already existing standardized database, \MF{} is several orders of magnitude faster than comparable optimization-based approaches for material model characterization from full-field measurements.
\CHANGE{The method provides parameter estimates that are close to the optimum identified by optimization-based approaches, while requiring only a fraction of the computational effort.
We also demonstrate that the models identified through \MF{} can serve as high-quality initial guesses for optimization-based methods.}
The database generated in this work can be used to \CHANGE{infer} constitutive models for other unseen materials, always following the testing guidelines and standardized sample geometry in our consciously designed experimental protocol.

\end{abstract}


\end{frontmatter}





\begin{figure}[ht]
    \centering
    \includegraphics[width=0.9\linewidth]{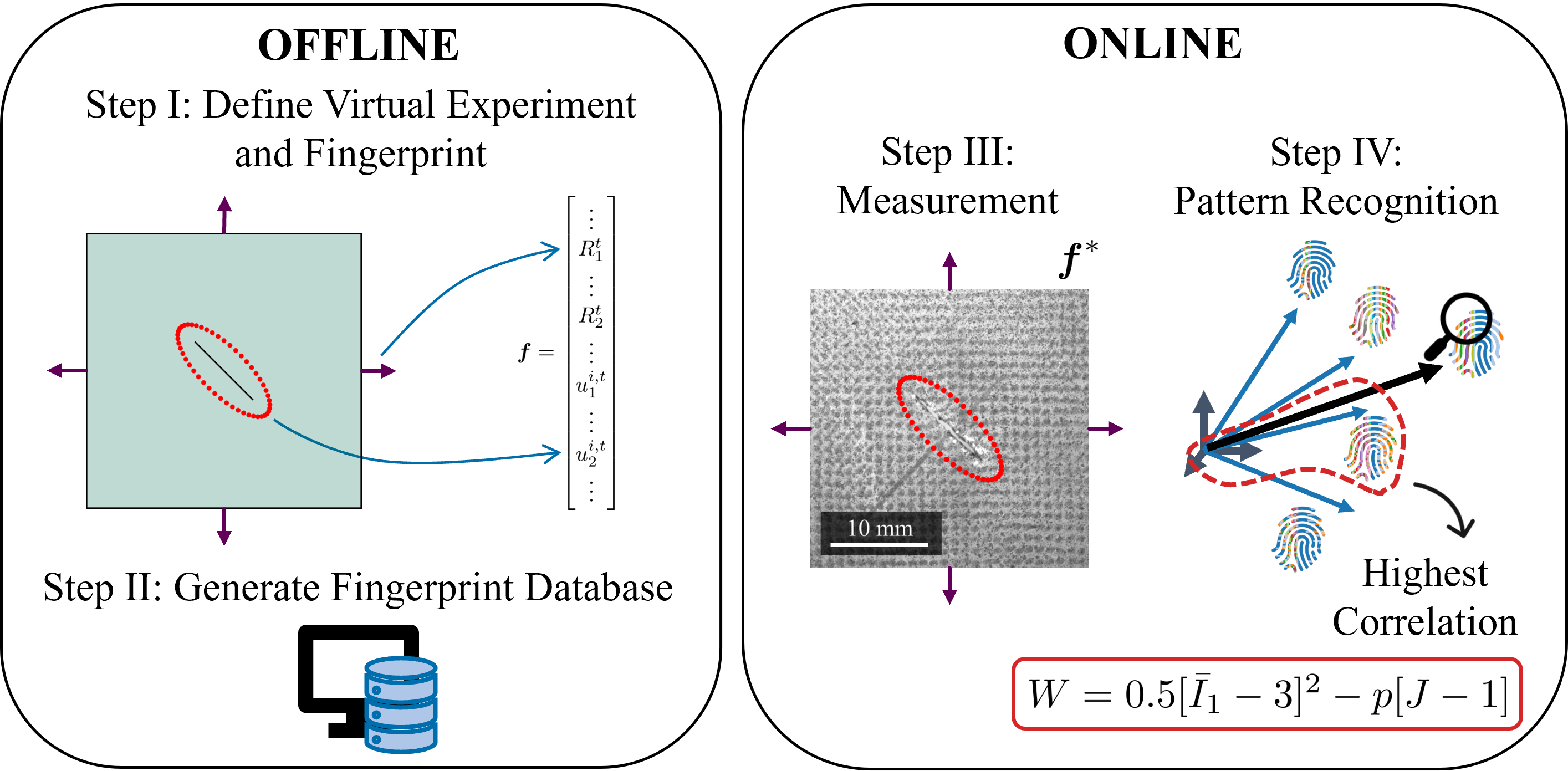}
    \caption{Schematic illustration of the unsupervised Material Fingerprinting workflow.
    In Step I, a standardized experiment and sample geometry is designed and the fingerprint vector is defined. In the unsupervised setting, the fingerprint incorporates both global reaction forces and local displacement measurements.
    In Step II, a database of fingerprints corresponding to different material models and parameter combinations is generated using finite element simulations. Once generated offline, this database can be reused indefinitely during the online stage.
    In Step III, the fingerprint of an unseen material is obtained experimentally following the standardized testing protocol and sample geometry.
    In Step IV, a pattern recognition algorithm is employed to \CHANGE{identify} the best‐matching fingerprint in the database and, consequently, the corresponding material model.
    }
    \label{fig:graphical_abstract}
\end{figure}


\section{Introduction}
\label{sec:introduction}

“\textit{Can Material Fingerprinting enable ultra-fast and robust material model discovery?}” This question has guided prior work on material model discovery \citep{flaschel_material_2026}. It now extends naturally to: “\textit{Can unsupervised Material Fingerprinting robustly uncover hyperelastic behavior directly from real full-field experimental data?}” 

Material modeling is fundamental to accurately capturing the constitutive response of a structure. While the identification of an appropriate strain energy density function was traditionally a trial-and-error process, recent advances in the material modeling community have led to a broad spectrum of methods for determining accurate constitutive descriptions and their parameters.
In contrast to tests on simply shaped specimens with homogeneous deformation fields, modern material characterization methods deliberately introduce controlled heterogeneity \CHANGE{into the deformation field}. The results are insightful, full-field experimental measurements \citep{avril_overview_2008,pierron_towards_2020,romer_reduced_2025}. As these methods rely on indirect, unlabeled data consisting of local displacements and global reaction forces, in contrast to direct, labeled data consisting of stress-strain data pairs, we will refer to such methods as "unsupervised" in machine learning jargon.

Among these unsupervised methods, the Finite Element Model Updating (FEMU) approach is widely used for calibrating the parameters of constitutive models (cf. the seminal work in \cite{Collins1974} or more recent implementations in \cite{wiesheier_versatile_2024} and the framework DAVIS for data-adaptive Generalized Standard Materials in \cite{wiesheier_data-adaptive_2026}). FEMU consists of solving the boundary value problem with the Finite Element Method (FEM) to reproduce the experimental setup, while the material parameters are iteratively optimized to minimize a residual, typically defined from the mismatch between experimental and computational displacement fields and reaction forces. Its main strengths are its versatility and robustness, as it can provide a material law for virtually any boundary value problem that can be represented in a forward finite element simulation. However, FEMU can be computationally expensive and may exhibit issues in parameter identification due to the potential existence of multiple local minima in the optimization landscape. 

Another commonly adopted identification technique is to calibrate the material model parameters so that the residuals of the governing weak formulation are minimized. This strategy has been proposed as the Virtual Fields Method (VFM) \citep{grediac_principle_1989,Grdiac1990,Grdiac2006} and the Equilibrium Gap Method (EGM) \citep{claire_finite_2004}.
An advancement of this strategy was subsequently used to automatically discover the optimal functional form of the material model through the EUCLID method (Efficient Unsupervised Constitutive Law Identification and Discovery) \citep{flaschel_unsupervised_2021,flaschel_discovering_2022,abbasi_discovery_2026}.
\CHANGE{The VFM can further be used to calibrate multiple candidate models and evaluate their respective fitting accuracies \citep{peshave_metrics_2024}}.
These methods introduce experimentally measured displacement fields and reaction forces directly into the weak form of the governing equations and solve for the material parameters by enforcing equality (or near equality) between the internal and external virtual work. This strategy is computationally efficient because it does not require repeated solutions of boundary value problems. However, these approaches can be sensitive to noise, as the measured displacement fields must be differentiated to obtain strains, and they may lack robustness when the input displacement fields are incomplete.

Traditionally, material characterization relies on selecting a constitutive model a priori and identifying its parameters by fitting the model response to experimental observations. While effective in many cases, this paradigm has proven insufficient for capturing increasingly complex material behaviors. This limitation has spurred the development of data-driven and machine learning-based approaches to material characterization. In these methods, the constitutive response may be represented by flexible black-box models, including neural networks \citep{ghaboussi_knowledgebased_1991,asad_mechanics-informed_2022,klein_polyconvex_2022,thakolkaran_nn-euclid_2022,klein_polyconvex_2025,masi_evolution_2023,rosenkranz_comparative_2023,linden_neural_2023,linka_new_2023,holthusen_theory_2024,kalina_neural_2025,flaschel_convex_2025,geuken_input_2025}, spline-based representations \citep{sussman_model_2009,wiesheier_versatile_2024,Dal2023,wiesheier_data-adaptive_2026,Moreno-Mateos2026}, Gaussian processes \citep{frankel_tensor_2020,fuhg_local_2022}, neural ordinary differential equations \citep{tac_data-driven_2022}, or parameterized non-smooth functions \citep{bleyer_learning_2025}. More recent approaches leverage large language models for the generation of constitutive models \citep{tacke_constitutive_2025}. Related approaches dispense with an explicit constitutive model altogether and instead perform simulations directly guided by experimental data \citep{kirchdoerfer_data-driven_2016,ibanez_data-driven_2017}. However, because purely data-driven and model-free strategies often lack physical interpretability, a complementary research direction has emerged that seeks to automatically infer interpretable constitutive models from data using symbolic or sparse regression techniques \citep{schoenauer_evolutionary_1996,flaschel_unsupervised_2021,flaschel_automated_2023,flaschel_non-smooth_2025,abdusalamov_automatic_2023,linka_new_2023,linka_best--class_2024,urreaquintero_automated_2026}. Rather than tuning parameters within a prescribed model class, these methods aim to uncover both the functional form of the material law and its associated parameters directly from experimental observations.
For a more comprehensive review of data-driven methods for constitutive modeling, we refer to \cite{fuhg_review_2024}.
However, a key limitation of existing data-driven approaches for material model learning and discovery is their reliance on optimization procedures, which are often computationally expensive and provide no guarantee of converging to the global optimum.

Very recently, we introduced Material Fingerprinting\footnote{
The term Material Fingerprinting is inspired by the Magnetic Resonance Fingerprinting (MRF) method by \cite{ma_magnetic_2013}, who proposed an inverse lookup table-driven strategy for quantitative magnetic resonance imaging, see also \cite{mcgivney_svd_2014,davies_compressed_2014,dong_datadriven_2025}. Similar strategies have been proposed in the field of rheology by \cite{rouze_characterization_2018, trutna_robust_2019, trutna_viscoelastic_2020,trutna_measurement_2020}. We also note that the terms Fingerprinting and Material Fingerprinting have been used outside the domain of mechanics \citep{spannaus_materials_2021,kuban_similarity_2022,filip_material_2024,jaafreh_introducing_2025}.} as a fast, robust, and easy-to-implement approach for mechanical material characterization \citep{flaschel_material_2026}, see \cref{fig:graphical_abstract}.
The method leverages a precomputed database in combination with a pattern recognition framework to determine the material model of a material.
The central premise is that each material exhibits a fingerprint that characterizes its mechanical response in a predefined experiment. These fingerprints are generated offline through forward finite-element simulations of a prescribed boundary value problem. In a subsequent online phase, a pattern recognition algorithm identifies the database entry that best matches the experimentally measured fingerprint.
This strategy offers several key advantages: it achieves exceptional computational efficiency, with the online identification stage completed within seconds. Moreover, Material Fingerprinting circumvents the need to solve a \CHANGE{continuous optimization problem}, which is often computationally expensive and prone to convergence issues, particularly in the presence of non-convex objective functions or nonlinear constraints. \CHANGE{In contrast to approaches that calibrate the parameters of a preselected material model, \MF{} automatically discovers an interpretable and suitable functional form of the constitutive model from a predefined set of models.} By restricting the database generation to physically admissible material models, the identified models are guaranteed to satisfy fundamental physical principles such as objectivity and thermodynamic consistency. Finally, the Material Fingerprinting framework is highly versatile and can be applied to a wide range of experimental setups and material behaviors. The method has been numerically verified for synthetically generated data  by \cite{flaschel_material_2026}. A first experimental investigation by \cite{martonova_material_2026} validated the approach using tests with homogeneous strain fields. Building on this work, a pip-installable, open-source Python package for \MF{}, tailored to experiments with homogeneous strain fields, has been released \citep{flaschel_python_2025}. \CHANGE{A recent adaptive implementation of the \MF{} framework has further demonstrated its ability to discover linear combinations of candidate models \citep{flaschel_adaptive_2026}.}
\CHANGE{Most of the aforementioned works on Material Fingerprinting focus on experiments involving homogeneous strain fields. However, the concept of Material Fingerprinting is equally applicable, in an unsupervised manner, to experiments with heterogeneous strain fields, which generally provide richer information from a single experimental test. For experiments with heterogeneous strain fields, inverse identification methods such as FEMU require repeated forward simulations and are therefore computationally demanding. Therefore, shifting this computational effort to an offline stage, where a reusable database is generated in the spirit of Material Fingerprinting, can significantly accelerate the subsequent identification process. A first proof-of-concept of unsupervised Material Fingerprinting applied to synthetically generated data with Gaussian-distributed noise was presented by \cite{flaschel_material_2026}. However, the application of \MF{} to real experimental tests involving heterogeneous strain fields remains an open question, which we address in the present contribution.}

In this contribution, we present the first application of unsupervised Material Fingerprinting to real experimental data consisting of full field measurements and reaction forces for the \CHANGE{inference} of hyperelastic behavior in five distinct materials. A database of material fingerprints associated with incompressible hyperelastic constitutive models is generated in an offline phase using forward finite element simulations of a predefined biaxial test. The considered specimen geometry contains a central cut, which induces heterogeneous strain fields. Subsequently, in an online phase, a pattern recognition algorithm identifies the database fingerprint that best matches the experimentally measured fingerprint. \CHANGETWO{In this context, we demonstrate that a precomputed database can be used for Material Fingerprinting even when the temporal and spatial resolutions of the simulations used during database generation do not exactly match those of the experimental measurements.} The proposed methodology is \CHANGE{demonstrated} using experimental data for three different grades of Elastosil, as well as Sylgard and VHB tape, which were previously acquired in the experimental study of \cite{moreno-mateos_biaxial_2025}.
\CHANGE{On a side note, we propose Material Fingerprinting as a first step to generate not strictly optimal, but ultra-fast estimates of material parameters that can be used as effective initial values for a classical (continuous) optimization to increase accuracy while keeping the computational effort to a minimum.}
The modus operandi of Material Fingerprinting is first demonstrated using a Cosine similarity based pattern recognition scheme introduced in earlier work \citep{flaschel_material_2026}. This similarity measure emphasizes the shape of the fingerprints while disregarding their absolute magnitudes. In the unsupervised setting considered here, each fingerprint incorporates multiple physical quantities, namely reaction forces and displacement fields, which makes it essential to retain information related to both shape and magnitude. For this reason, we introduce a novel similarity measure for pattern recognition based on a Euclidean metric. This metric has not previously been investigated within the Material Fingerprinting framework and is particularly well suited for the present application, as it preserves variations in magnitude as well as the geometric structure of the fingerprints.

In this study, the \MF{} database is generated according to a standardized experimental protocol and sample geometry that are designed to be reproducible across a wide range of materials in laboratories. It is important to note that any database used for Material Fingerprinting is inherently tied to a fixed experimental configuration, including both specimen geometry and loading conditions. Consequently, any modification to the experimental setup necessitates the generation of a new database. While this requirement represents a limitation of the approach, reliance on standardized testing protocols for material characterization is both reasonable and well established in the literature.
Analogous to existing technical standards, such as those defined by DIN, EN, ISO, or ASTM, the Material Fingerprinting framework envisions databases that are accompanied by clearly defined experimental guidelines specifying the required sample geometry and loading conditions to ensure consistency with the fingerprints database. The key idea is that the database is generated only once, and then it can be reused many times during the online phase across different laboratories and for different materials, following a consistent, standardized, and easy testing guideline. Here, the database generated in the offline phase is to be interpreted as a lookup table. Furthermore, supplementary files for the three-dimensional printing of molds required for in-house specimen fabrication could be provided as part of a dedicated Material Fingerprinting handbook. Once such experimental guidelines are established and a comprehensive database is generated and made publicly available, unsupervised material characterization via \MF{} becomes highly efficient and robust.

\section{\MF{}}
\label{sec:material_fingerprinting}

In this work, we focus on the unsupervised setting of \MF{}, in which local displacement data over the specimen surface and global reaction forces at the specimen boundary are available, but no local stress measurements are provided.
Because different materials produce distinct displacement and force responses, a vector containing both quantities can be interpreted as the material’s fingerprint.

To enable model \CHANGE{identification}, we first construct a database of simulated fingerprints for a wide range of material models and parameter combinations using finite element simulations.
Combined with an appropriate pattern recognition algorithm, this database can then be queried repeatedly to identify suitable material models and their parameters for previously unseen materials. The main premise is that the database is generated only once and then reused repeatedly during the ultra-fast online phase across different laboratories and for different materials, following a testing guideline with a standardized experiment and a sample geometry.

In the following, we describe the standardized experiment, the definition of the fingerprints, the database generation procedure, and the pattern recognition framework underlying Material Fingerprinting, which we later apply to identify material models from experimentally measured fingerprints.

\subsection{Standardized experiment}
\label{sec:standardized_experiment}

\begin{figure}[ht]
    \centering
    \includegraphics[width=0.4\linewidth]{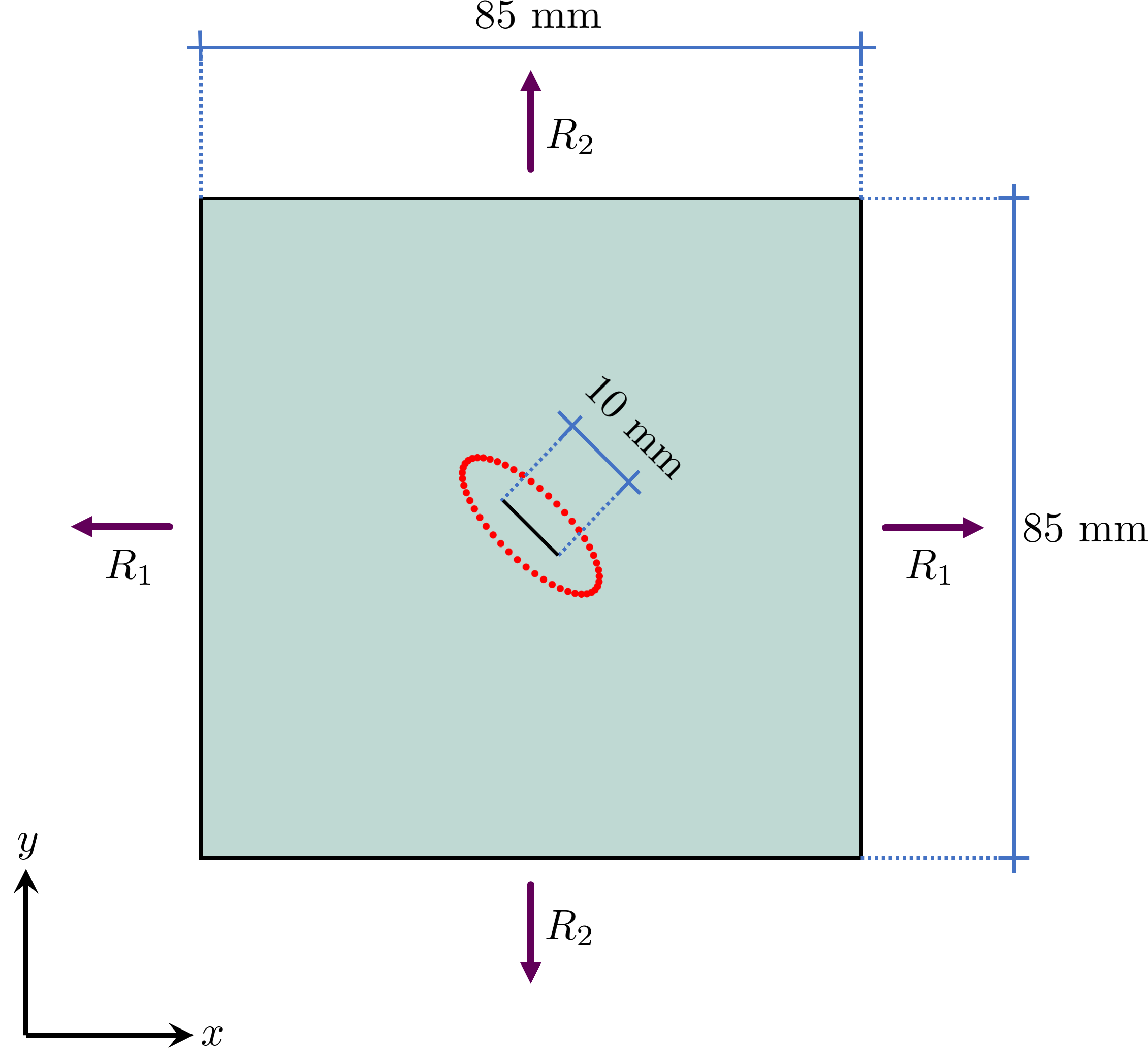}
    \caption{Illustration of the standardized experimental setup. The specimen is under biaxial tension. The global reaction forces can be measured via load cells and the displacements at the red markers can be measured using full-field measurement or local point tracking techniques.}
    \label{fig:geometry}
\end{figure}

As described previously, our experimental tests follow a standardized procedure: all experimentally tested specimens share the same geometry, and identical loading conditions are applied.
For the database generation, we consider a biaxial test of a thin specimen with a central notch
\footnote{\CHANGE{
We note that alternative specimen geometries can also be employed to induce heterogeneous deformation fields. For example, specimens with circular or ellipsoidal holes are commonly used in the material discovery community \citep{abbasi_discovery_2026}. In the present work, however, we adopt a specimen with a central notch primarily because of its ease of fabrication. Introducing a notch of prescribed length can be achieved with greater geometric accuracy and reproducibility than manufacturing an ellipsoidal cutout of prescribed dimensions.
Moreover, the chosen specimen geometry is consistent with the experimental data extracted from \cite{moreno-mateos_biaxial_2025}. While the experiments reported in \cite{moreno-mateos_biaxial_2025} were conducted up to failure, the present study is restricted to load steps preceding the onset of damage localization.
}}
to introduce heterogeneity in the strain field, as illustrated in \cref{fig:geometry}.
The side length of the square specimen measures \qty{85}{\milli \meter}, and a thickness of \qty{2}{\milli \meter} is assumed.
As discussed later, the specimen is assumed to be sufficiently thin such that the reaction forces scale linearly with specimen thickness for fixed applied displacements.
The notch is oriented at \qty{45}{\degree} with respect to the loading axes and has a length of \qty{10}{\milli \meter}.
We apply an equibiaxial displacement of \qty{29.75}{\milli \meter} over 35 equidistant steps, at which reaction forces and displacements can be recorded.
We chose this predefined experimental protocol such that it can be reproduced experimentally for different materials across laboratories. 

\subsection{Fingerprint definition}

In the following, we define the material fingerprint tailored to the aforementioned standardized experimental setup.
Specifically, a fingerprint is a real-valued vector $\bff$ that contains both the reaction forces and displacements resulting from the applied deformation. Therefore, a fingerprint encodes the characteristic mechanical response of the material.
During the experiment, the reaction forces measured for each axis at the specimen boundary across a predefined set of load steps are collected in a vector $\bff_R \in \Rset^{2 \, n_t}$, where $n_t$ denotes the number of load steps.
Likewise, the displacements measured at a predefined set of surface points are assembled into a vector $\bff_u \in \Rset^{2 \, n_u \, n_t}$, where $n_u$ is the number of surface points contributing to the fingerprint.
\cref{fig:geometry} illustrates the surface points selected for constructing the fingerprint. Because the deformation field exhibits the greatest variation near the central crack, we place the measurement points along an ellipse surrounding this region.
At the same time, full-field measurements tend to be less reliable near specimen boundaries, so the ellipse is chosen with a sufficient radius to maintain a buffer zone between the notch and the measurement points. Note that alternative sampling distributions are possible, provided they maintain a prudent distance from the contour. The ellipsoidal layout, however, keeps the fingerprint compact while still capturing the field heterogeneity near the notch.
The reaction forces and displacements are concatenated into the fingerprint vector $\bff = [\bff_R ; \bff_u]\in\Rset^{n_f}$ with $n_f = 2 \, n_t + 2 \, n_u \, n_t$.
Here, we note that the order in which the measurements appear in the fingerprint vector is irrelevant, provided that this ordering is applied consistently throughout all simulations and experiments.

\subsection{Database generation}
\label{sec:database_generation}

To generate a database of material fingerprints, we virtually reproduce the standardized experimental setup described in \cref{sec:standardized_experiment} and simulate its response using three-dimensional finite element computations for a predefined set of material models and parameter combinations.
In this work, we focus on incompressible and isotropic materials for which the material behavior is characterized by the strain energy density function
\begin{equation}
    W(\bfF;\bftheta,\bfalpha) = \bar{W}(\bfF;\bftheta,\bfalpha) - p[J-1],
\end{equation}
where $\bfF=\nabla_0 \, \bfu + \bfI$ is the deformation gradient of the displacement field $\bfu$, with $\bfI$ the second-order identity tensor, $p$ is a Lagrange multiplier enforcing the incompressibility constraint $J=\det(\bfF)=1$, $\bftheta$ is a vector of homogeneity material parameters while $\bfalpha$ is a vector of non-homogeneity material parameters (the difference between the last two will be explained in \cref{sec:homogeneity}).
For isotropic materials, the isochoric strain energy density $\bar{W}$ may be a function of the invariants $\bar I_1,\bar I_2$ or the principal stretches $\bar \lambda_1,\bar \lambda_2,\bar \lambda_3$.
Specifically, we have $\bar I_1 = J^{-2/3}I_1$ and $\bar I_2=J^{-4/3}I_2$, where $I_1=\text{tr}(\bfC)$ and $I_2=\frac{1}{2}[\text{tr}(\bfC)^2 - \text{tr}(\bfC^2)]$ are the first and second invariants of the right Cauchy-Green tensor $\bfC=\bfF^T \cdot \bfF$, and $\bar \lambda_1,\bar \lambda_2,\bar \lambda_3$ are the square roots of the eigenvalues of $\bar \bfC = J^{-\frac{2}{3}} \bfC$.

We consider seven incompressible hyperelastic material models: the Carroll model \citep{carroll_strain_2011}, the Lopez-Pamies model \citep{lopez-pamies_new_2009}, the Mooney-Rivlin model \citep{rivlin_large_1950}, the Neo-Hookean model \citep{treloar_stress-strain_1944}, the generalized Neo-Hookean model \citep{stephenson_equilibrium_1982,geubelle_finite_1994}, the Ogden model \citep{ogden_large_1972}, and the Yeoh model \citep{yeoh_characterization_1990}.
\CHANGE{For all models, the isochoric strain energy density $\bar W$ is formulated in terms of the isochoric invariants $\bar I_1, \bar I_2$ or the isochoric principal stretches $\bar \lambda_i$. By construction, the models satisfy thermodynamic consistency, material frame indifference, isotropy, and a stress-free reference configuration.}
For each model, we perform multiple simulations across a range of parameter values and store the resulting fingerprints, along with the corresponding model metadata, in the database.
\cref{tab:database_unsupervised} summarizes the material models, their strain energy density functions, and the parameter ranges used during database generation.
\CHANGE{We note that we sample the homogeneity parameters such that no homogeneity parameter is more than an order of magnitude smaller than all other homogeneity parameters in the model. The rationale behind this choice is that we consider modeling terms to be practically irrelevant relative to the other modeling terms if their corresponding homogeneity parameter is more than ten times smaller. The non-homogeneity parameters, however, have different mathematical interpretations across the considered material models. For example, they may appear either as multiplicative factors or as exponents. Therefore, the sampling ranges are selected individually for each non-homogeneity parameter to ensure that they cover physically meaningful and representative parameter regimes.}
\CHANGE{At this point, we also note that, due to the discretization of the parameter space, \MF{} should be regarded as a method for identifying approximate material parameters rather than strictly optimal values over the continuous parameter domain. As we will demonstrate later, however, the resulting loss in accuracy is negligible for practical applications.}
Each simulation results in a fingerprint vector $\bff^{(i)}$ with $i=1,\dots,n_d$, where $n_d$ denotes the number of fingerprints in the database.
The finite element simulations are implemented in the open-source finite element platform FEniCSx. Details about the implementation are provided in \cref{sec:FE_simulations}. 

\begin{table}[h!]
\caption{
Material models and parameter combinations considered during database generation.
}
\label{tab:database_unsupervised}
\centering
\resizebox{\textwidth}{!}{
\begin{tabular}{|l|l|l|r|}
\hline
Models\vphantom{$\int_a^b$} & Strain energy density $\bar{W}$ & \multicolumn{1}{c|}{Parameters ranges} & \# Fingerprints \\ \hline


\multicolumn{1}{|l|}{Carroll\vphantom{$\int_a^b$}} & $\theta_1 \bar I_1 + \theta_2 \bar I_1^4 + \theta_3 \sqrt{\bar I_2}$ & $\theta_1 = 1.0$, $\theta_2 \in [0.1, 10.0]$, $\theta_3 \in [0.1, 10.0]$ & 100\\ 



\multicolumn{1}{|l|}{Lopez-Pamies\vphantom{$\int_a^b$}} & $\theta_4 [\bar I_1^{\alpha_1} - 3^{\alpha_1}]$ & $\theta_4 = 1.0$, $\alpha_1 \in [0.01, 10.0]$ & 100\\ 

\multicolumn{1}{|l|}{Mooney-Rivlin\vphantom{$\int_a^b$}} & $\theta_5 [\bar I_1 - 3] + \theta_6[\bar I_2 - 3]$ & $\theta_5 = 1.0$, $\theta_6 \in [0.1, 10.0]$ & 100\\

\multicolumn{1}{|l|}{Neo-Hookean\vphantom{$\int_a^b$}} & $\theta_5 [\bar I_1 - 3]$ & $\theta_5 = 1.0$ & 1\\ 

\multicolumn{1}{|l|}{Gen. Neo-Hookean\vphantom{$\int_a^b$}} & $\theta_7 \left[\left[1+\alpha_2[\bar I_1 - 3]\right]^{\alpha_3} - 1\right]$ & $\theta_7 = 1.0$, $\alpha_2 \in [0.01, 10.0]$, $\alpha_3 \in [0.5, 10.0]$ & 400\\ 

\multicolumn{1}{|l|}{Ogden\vphantom{$\int_a^b$}} & $\theta_8 [\bar \lambda_1^{\alpha_4} + \bar \lambda_2^{\alpha_4} + \bar \lambda_3^{\alpha_4} - 3]$ & $\theta_8 = 1.0$, $\alpha_4 \in [1.0, 10.0]$ & 100 \\ 

\multicolumn{1}{|l|}{Yeoh\vphantom{$\int_a^b$}} & $\theta_5 [\bar I_1 - 3] + \theta_9 [\bar I_1 - 3]^2 + \theta_{10} [\bar I_1 - 3]^3$ & $\theta_5 = 1.0$, $\theta_9 \in [0.1, 10.0]$, $\theta_{10} \in [0.1, 10.0]$ & 100 \\ 

\hline
\multicolumn{3}{|c|}{\vphantom{$\int_a^b$}} & $n_d = 901$ \\
\hline
\end{tabular}
}
\end{table}

For the database generation, the number of load steps $n_t$ is chosen to be sufficiently large to span the entire elastic regime up to the onset of crack propagation for all experiments. Selecting a sufficiently large $n_t$ ensures that the database remains applicable to future experiments, even when the exact load step at which crack propagation will occur cannot be anticipated.
We note that incompressible finite element simulations may suffer from convergence issues, particularly at larger deformations or for material models exhibiting strong nonlinearities. During database generation, if a simulation fails to converge at a given load step, we set all fingerprint entries for that load step and all subsequent load steps to zero. As a result, the fingerprint is no longer usable across the full load range but remains valid for experimental data restricted to the lower load regime.

\CHANGE{The generation of the database is computationally demanding and time-consuming. However, this cost is incurred only once during the offline stage. The resulting database can then be reused during the ultra-fast online phase across different laboratories and for a wide range of materials, provided that a standardized testing protocol, including a prescribed experimental setup and specimen geometry, is followed.
For the database considered in this work, the cumulative computational effort amounts to approximately 46 CPU hours. We report this quantity for completeness. Within the \MF{} framework, this one-time offline investment enables repeated online identification without requiring additional finite element simulations in the online phase.}

\subsection{Homogeneity property}
\label{sec:homogeneity}

The material parameters directly influence the resultant forces and displacements in the finite element simulations.
We distinguish between so-called homogeneity parameters $\bftheta$ and non-homogeneity parameters $\bfalpha$ \citep{flaschel_material_2026}, such that the isochoric strain energy density functions can be written as a linear combination of the homogeneity parameters and functions $\bfQ$ of the deformation and the non-homogeneity parameters
\begin{equation}
\label{eq:homogeneity}
    \bar{W}(\bfF;\bftheta,\bfalpha) = \bftheta \cdot \bfQ(\bfF;\bfalpha).
\end{equation}
When the finite element analysis is performed under pure displacement control, multiplying the strain energy density by a factor scales the reaction forces proportionally, while the displacement field remains unaffected.
Specifically, the forces obey the homogeneity relation $\bff^{(i)}_R(a \bftheta, \bfalpha) = a \bff^{(i)}_R(\bftheta, \bfalpha), \ \forall a \in \Rset$, whereas the displacements remain unchanged under such scaling $\bff^{(i)}_u(a \bftheta, \bfalpha) = \bff^{(i)}_u(\bftheta, \bfalpha), \ \forall a \in \Rset$.
These relations are useful because, after the pattern recognition algorithm identifies the material model, they allow the homogeneity parameters $\bftheta$ to be properly scaled to match the observed forces.
We note that the functional form in \cref{eq:homogeneity} imposes no restrictions on hyperelastic material models because the functions $\bfQ$ may depend nonlinearly on the parameters $\bfalpha$, and all models in the literature fall within this form.

\CHANGE{The distinction between homogeneous and non-homogeneous parameters in Material Fingerprinting is a powerful concept that is consistent with related approaches in the literature. For example, \cite{Perotti2017} investigated constitutive formulations in which the strain energy density is expressed as a linear combination of nonlinear polyconvex features. The same authors proposed efficient calibration strategies for parameters entering the material model linearly and nonlinearly \citep{Motevalli2023}. \cite{flaschel_automated_2023-1} developed an automated strategy for material model discovery in the presence of both linear and nonlinear model parameters.
}

\subsection{Pattern recognition algorithm with Cosine similarity}

After generating the database, it can be leveraged repeatedly for material model \CHANGE{identification}.
To this end, the standardized experiment must be conducted for the material under consideration, and the material's fingerprint must be measured.
In the following, we denote the experimentally measured fingerprint by $\bff^*$.
We note, however, that it is not always possible to acquire data for all load steps $n_t$ experimentally, as the material may fail before reaching the final load step.
Thus, it might not always be possible to acquire the entire fingerprint.
In this case, a reduced fingerprint $\hat\bff^*\in\Rset^{\hat{n}_f}$ with $\hat{n}_f < n_f$ is measured and used for Material Fingerprinting by loading only the corresponding reduced fingerprints $\hat\bff^{(i)}$ from the database.

Given an experimentally measured fingerprint $\hat\bff^*$, the Material Fingerprinting method \CHANGE{identifies} a suitable material model by searching the database for the best matching fingerprint $\hat\bff^{(i)}$.
To this end, the measured fingerprint and the fingerprints in the database are normalized
\begin{equation}
\label{eq:normalization}
    \bar\bff^*_R = \frac{\hat\bff^*_R}{\|\hat\bff^*_R\|}, \ 
    \bar\bff^*_u = \frac{\hat\bff^*_u}{\|\hat\bff^*_u\|}, \ 
    \bar\bff^{(i)}_R = \frac{\hat\bff^{(i)}_R}{\|\hat\bff^{(i)}_R\|}, \
    \bar\bff^{(i)}_u = \frac{\hat\bff^{(i)}_u}{\|\hat\bff^{(i)}_u\|}. \ 
\end{equation}

Due to normalization, the inner products of the fingerprints are equal to the cosines of the angles between the fingerprints
\begin{equation}
\label{eq:angles}
    \cos(\beta_R^{(i)}) = \bar\bff^{(i)}_R \cdot \bar\bff^*_R, \
    \cos(\beta_u^{(i)}) = \bar\bff^{(i)}_u \cdot \bar\bff^*_u, \
\end{equation}
where $\beta_R^{(i)}$ is the angle between $\bar\bff^{(i)}_R$ and $\bar\bff^*_R$, and $\beta_u^{(i)}$ is the angle between $\bar\bff^{(i)}_u$ and $\bar\bff^*_u$.
Hence, the inner products are also called Cosine similarities. Because the inverse cosines $\cos^{-1}(\square)$ are monotonically decreasing functions for $\square\in[-1,1]$, the Cosine similarities are inversely proportional to the angles between the fingerprints.

We finally identify the fingerprint in the database with the highest agreement with the experimentally measured data. To this end, we seek the maximum of a weighted measure between the Cosine similarities of the reaction and displacement fingerprints
\be
\label{eq:pattern_recognition_unsupervised}
i_C^*
=
\arg\max_{i=1,\dots,n_d}
\left[
\frac{1}{n_{f_R}} \cos(\beta_R^{(i)})
+
\frac{1}{n_{f_u}} \cos(\beta_u^{(i)})
\right],
\ee
where $n_{f_R}$ and $n_{f_u}$ denote the dimensions of the fingerprints $\bar\bff^*_R$ and $\bar\bff^*_u$, respectively. \CHANGE{We note that the problem above is a discrete rather than a continuous optimization problem, also known as a nearest neighbor search problem. Unlike continuous optimization problems, finding the maximum over a finite set of discrete values is computationally inexpensive and guarantees identification of the global optimum.} \CHANGE{We further note that solving the pattern recognition problem described above does not require any forward finite element simulations, as these simulations have already been carried out during the offline stage. Therefore, the online identification step is computationally much less expensive than solving a continuous inverse optimization problem.}

The reader might note that the normalization in \cref{eq:normalization} serves to place the reaction force and displacement contributions on a comparable scale.
Both contribute equally to the recognition algorithm in \cref{eq:pattern_recognition_unsupervised}, ensuring that neither dominates the combined measure. In principle, we can adjust the relative impact of the reaction forces and displacements by introducing weights into the similarity measure.
However, our numerical experiments indicate that our chosen measure of similarity consistently yields appropriate and robust results.

We finally obtain the identified material model by rescaling the homogeneity parameters with the norm of the force measurements
\be
\label{eq:rescaling_unsupervised}
\bftheta^* = \|\hat\bff^*_R\| \, \bar\bftheta^{(i^*_C)} \quad \text{and} \quad \bfalpha^* = \bfalpha^{(i^*_C)}.
\ee

\subsection{Pattern recognition algorithm with Euclidean similarity}
The Cosine similarity-based measure introduced in the previous section enables fingerprint recognition by primarily emphasizing the shape of the fingerprints. However, since the reaction forces and displacements concatenated into a fingerprint are normalized (cf. \cref{eq:normalization}), the proportionality between forces and displacements is not preserved. This motivates an alternative similarity measure based on Euclidean distance. Specifically, we define a second similarity metric using the Euclidean norm of the difference between the database and experimental reaction forces and displacements as
\be
\label{eq:pattern_recognition_unsupervised_Euclidean}
i_E^*
=
\arg\max_{i=1,\dots,n_d}
\left[
-
\dfrac{\|\bar{\bff}^{(i)}_R - \bar{\bff}^{*}_R \|^2}
      {\|\bar{\bff}^{*}_R\|^2}
-
\dfrac{\|\hat{\bff}^{(i)}_u - \hat{\bff}^{*}_u \|^2}
      {\|\hat{\bff}^{*}_u \|^2}
\right].
\ee
The first term in \cref{eq:pattern_recognition_unsupervised_Euclidean} involves normalized force fingerprints. The use of the normalized vectors $\bar{\bff}^{(i)}_R$ and $\bar{\bff}^{*}_R$ with $\|\bar{\bff}^{(i)}_R\| = \|\bar{\bff}^{*}_R\| = 1$ is motivated by the homogeneity property of the strain energy density function. The homogeneity property described in \cref{sec:homogeneity} leads to a proportional scaling of the reaction forces under a uniform scaling of the strain energy density function. Thus, it is sufficient to compare the relative shapes of the force fingerprints during pattern recognition and to determine the correct magnitude of the strain energy density function in a subsequent postprocessing step.
The displacement field remains invariant under a uniform scaling of the strain energy density function. Thus, the second term of the Euclidean similarity measure involves the non-normalized displacement fingerprints $\hat{\bff}^{(i)}_u$ and $\hat{\bff}^{*}_u$. Here, the squared norm of the difference between the measured fingerprint and each fingerprint in the database is divided by the squared norm of the experimental fingerprint to obtain a nondimensional measure of similarity. The pattern recognition algorithm in \cref{eq:pattern_recognition_unsupervised_Euclidean} is reformulated as a maximization problem by changing the sign of the error norms such that the resulting term can be interpreted as a measure of similarity.

We emphasize that the quantity maximized in \cref{eq:pattern_recognition_unsupervised_Euclidean} is similar -- but not identical -- to the objective function typically minimized in parameter identification with the FEMU method. In FEMU, the goal is to minimize the absolute differences between measured and simulated reaction forces. In contrast, our approach minimizes the discrepancy between normalized reaction force fingerprint vectors. Consequently, we focus solely on the direction of the reaction force fingerprint vectors while disregarding their magnitudes. This is justified in the Material Fingerprinting context, as the correct force magnitudes can be recovered later during the rescaling of the homogeneity parameters. After identifying $i^*_E$, the rescaling of the homogeneity parameters is analogous to \cref{eq:rescaling_unsupervised}.

\section{Experimental data}
\label{sec:experimental_data}

\begin{figure}[ht]
    \centering
    \includegraphics[width=1\linewidth]{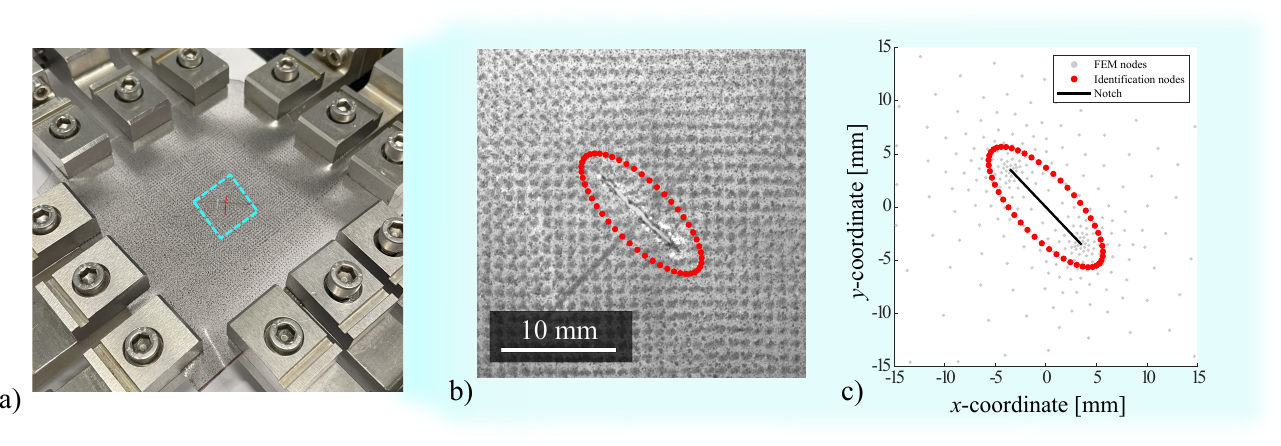}
    \caption{\textbf{Experimental fingerprint measurement.} (a) Experimental setup for the equibiaxial tests in \cite{moreno-mateos_biaxial_2025}. (b) Close-up view of the strain concentrator (notch) and its vicinity. The points used to construct the fingerprints are marked in red color. (c) Close-up view of computational counterpart of the notch and points used to construct the fingerprints in the database.}
    \label{fig:experiment}
\end{figure}

In this work, we \CHANGE{present the first experimental demonstration of unsupervised \MF{}} using the experimental data reported by the authors in \cite{moreno-mateos_biaxial_2025} under biaxial loading conditions. The campaign was carried out on standardized square samples made from five different soft materials (shear modulus below \qty{250}{\kilo\pascal}). Four of the materials were synthesized by curing two-component blends at elevated temperatures: Elastosil P7670 at three volume mixing ratios (2:1, 8:5, and 1:1, ranging from softest to stiffest) and Sylgard 184 at the standard 10:1 ratio. All samples had a thickness of \qty{2}{\milli \meter}, \CHANGE{except for VHB 4905 tape, which was} supplied by the manufacturer as pre-formed sheets of thickness \qty{0.5}{\milli\meter}. In addition, a notch in the middle introduces deformation heterogeneity. The notch is oriented at \qty{45}{\degree} with respect to the loading axes and has a length of \qty{10}{\milli \meter} (cf. \cref{fig:experiment}a-b). It is worth noting that the notch, acting as a strain concentrator, may induce inelastic effects in its immediate vicinity. However, these are minor prior to sample failure and diminish further away from the notch boundary.

Equibiaxial tests under monotonic loading were performed at a quasi-static displacement rate of \qty{0.85}{\milli\meter\per\second} along each axis, starting from an initial clamp separation of \qty{85}{\milli\meter}. Throughout the tests, 2D Digital Image Correlation (DIC) was applied to images captured during deformation. In this work, we restrict our analysis to the displacement fields measured at loading steps strictly preceding sample failure. The fingerprints are constructed as the geometric mean across four experimental repetitions. Details regarding the experimental fingerprint measurement are provided in \cref{sec:experimental_fingerprint_measurement}.

We note that the experimental dataset provides a sufficiently dense set of load steps such that we can extract the fingerprint information at the load steps predefined in \cref{sec:standardized_experiment}. The experimental displacement field is sampled at nodes lying on an elliptic contour, as illustrated in \cref{fig:experiment}b-c for both a representative sample and the finite element mesh used to generate the fingerprint database. This is possible due to the high spatial resolution of the experimental displacement fields obtained through DIC, which are originally computed on a square grid that is significantly denser than the set of points required for fingerprint construction.

Importantly, the experimental dataset is standardized with respect to both loading conditions and sample geometry. In this regard, it is noteworthy to mention that VHB tape samples have a smaller thickness due to fabrication constraints. To account for the geometric discrepancy relative to the other four materials, the reaction forces in the experimental dataset for VHB tape are scaled by a factor of four, reflecting that its thickness is one quarter of the thickness of the virtual sample used to generate the database. This scaling is justified by an independent \CHANGE{study} in which forward simulations were performed for both the database thickness and the reduced VHB tape thickness, demonstrating that the resulting force-displacement curves scale almost perfectly linearly with thickness and that the deformation fields barely vary.

\section{Results and discussion}

We apply \MF{} to the experimental data from the five materials described in \cref{sec:experimental_data}, using a single fingerprint database throughout the study, as discussed in \cref{sec:database_generation}.
We first present the results obtained using Cosine similarity for pattern recognition, and then discuss the results obtained with the newly proposed Euclidean similarity measure.
\CHANGE{We further assess the computational cost of \MF{} and compare it with that of optimization-based methods.
In addition, we present a representative study in which the parameter set identified by \MF{} is used as the initial guess for an optimization-based approach.}

\begin{figure}[!ht]
    \centering
    \begin{subfigure}[b]{0.4\textwidth}
        \centering
        \includegraphics[width=\textwidth]{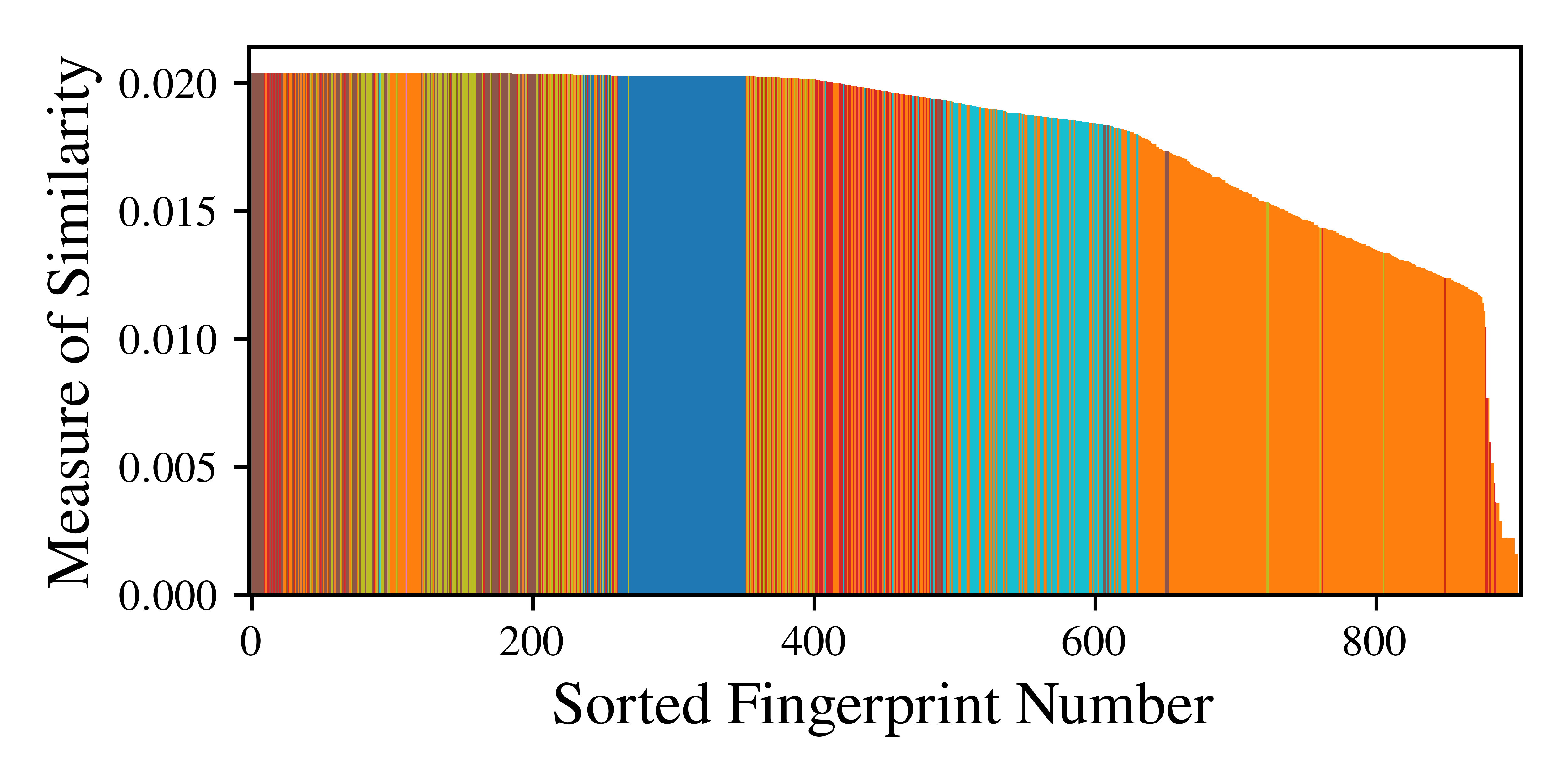}
        \caption{Elastosil 1:1.}
    \end{subfigure}
    \begin{subfigure}[b]{0.4\textwidth}
        \centering
        \includegraphics[width=\textwidth]{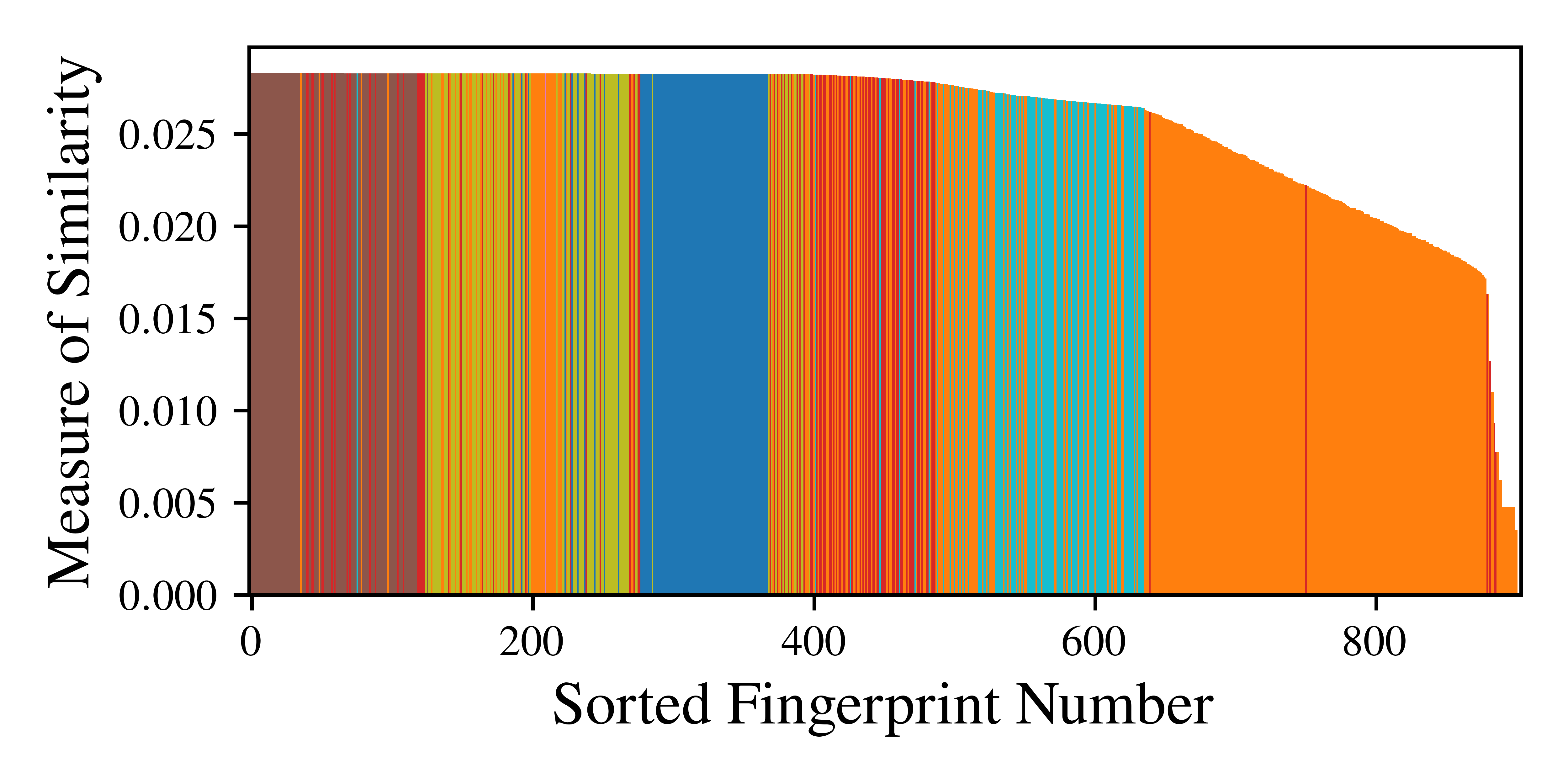}
        \caption{Elastosil 2:1.}
    \end{subfigure}
    \begin{subfigure}[b]{0.4\textwidth}
        \centering
        \includegraphics[width=\textwidth]{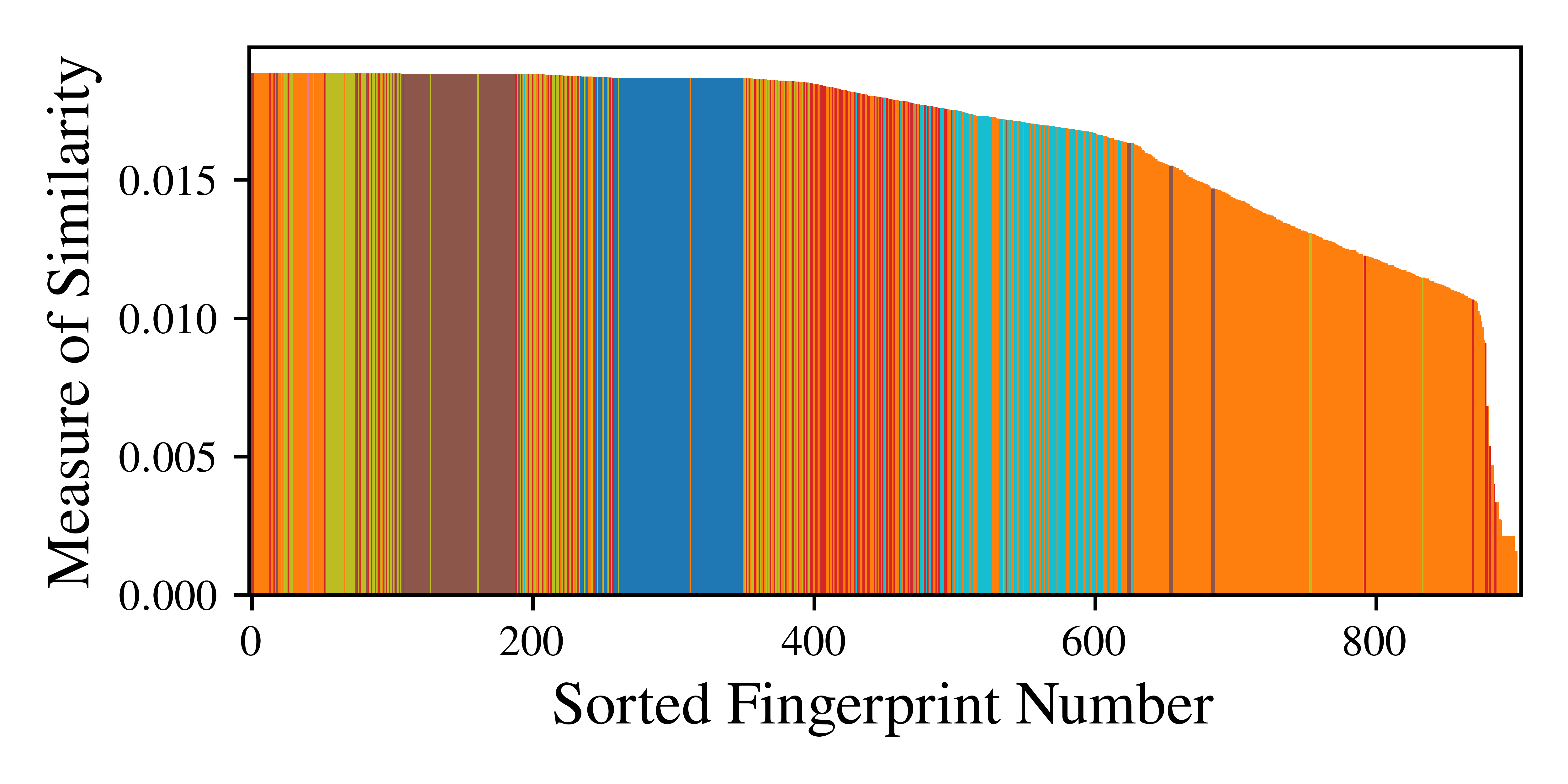}
        \caption{Elastosil 8:5.}
    \end{subfigure}
    \begin{subfigure}[b]{0.4\textwidth}
        \centering
        \includegraphics[width=\textwidth]{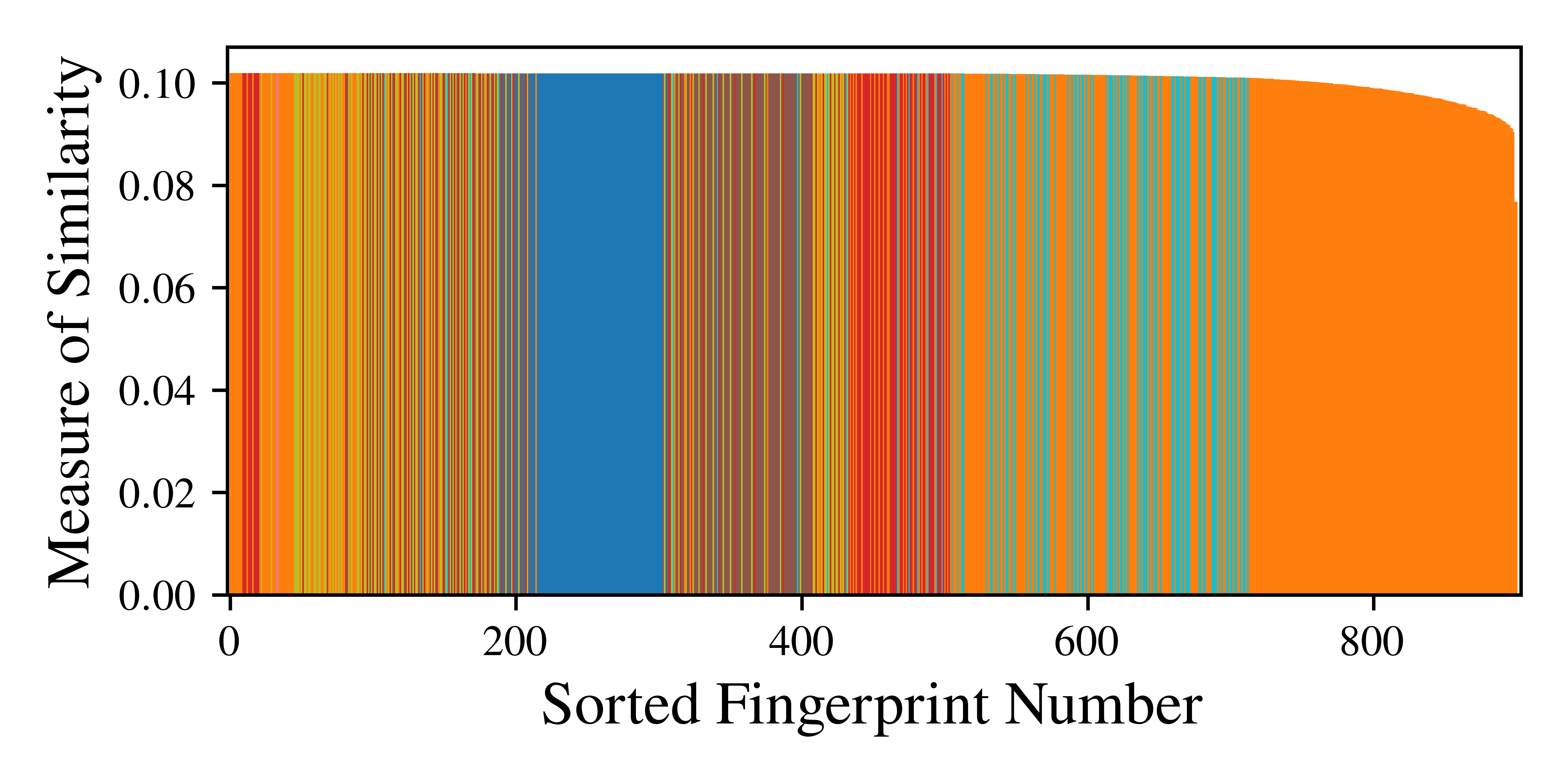}
        \caption{Sylgard.}
    \end{subfigure}
    \begin{subfigure}[b]{0.4\textwidth}
        \centering
        \includegraphics[width=\textwidth]{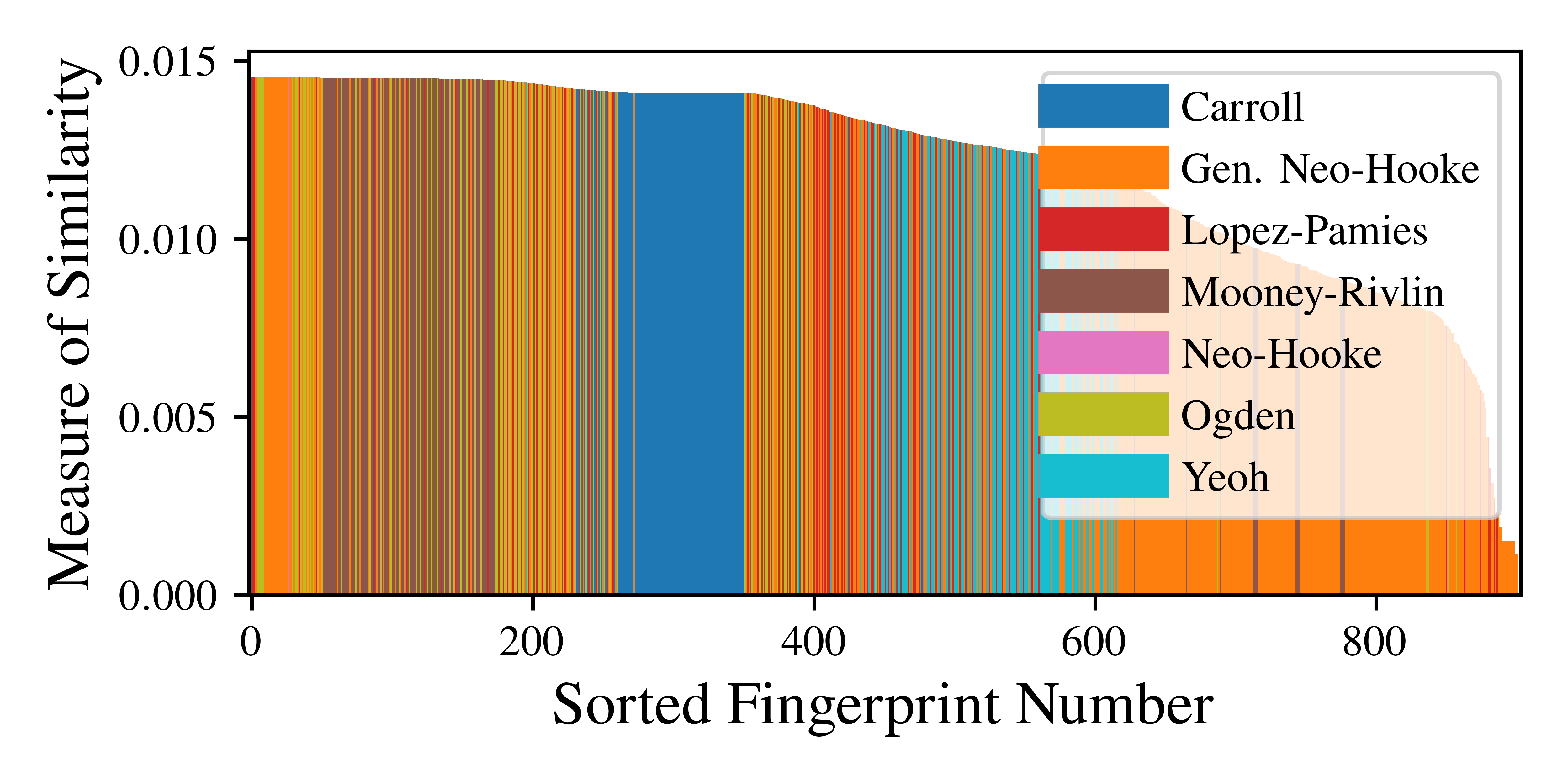}
        \caption{VHB tape.}
    \end{subfigure}
\caption{Illustration of the pattern recognition algorithm for Material Fingerprinting based on Cosine similarity. The plots displays the similarity measures for all fingerprints, sorted in descending order.}
\label{fig:cosine_similarity}
\end{figure}

\subsection{Cosine similarity}

To illustrate the inner workings of Material Fingerprinting, and in particular the pattern recognition based on Cosine similarity, we show in \cref{fig:cosine_similarity} the Cosine similarity computed for all fingerprints in the database.
These similarities are ordered by magnitude, such that the most suitable fingerprint appears on the left.
\CHANGE{These plots highlight that multiple fingerprints in the database can yield similar similarity values, indicating that several models and combinations of material parameters are comparably capable of describing the data.
This behavior is expected, as some models in the database can be viewed as generalizations of others, as is the case for the Neo-Hookean, generalized Neo-Hookean, and Ogden models.}
We observe that, particularly for Sylgard, many models exhibit similar Cosine similarity values.
This is reasonable because, as shown later, the material response of Sylgard is close to linear under the experimental conditions. \CHANGE{The experimental data are restricted to relatively small deformations, with strains remaining below \qty{10}{\%}.}

\begin{figure}[!ht]
    \centering
    \begin{subfigure}[b]{0.7\textwidth}
        \centering
        \includegraphics[width=\textwidth]{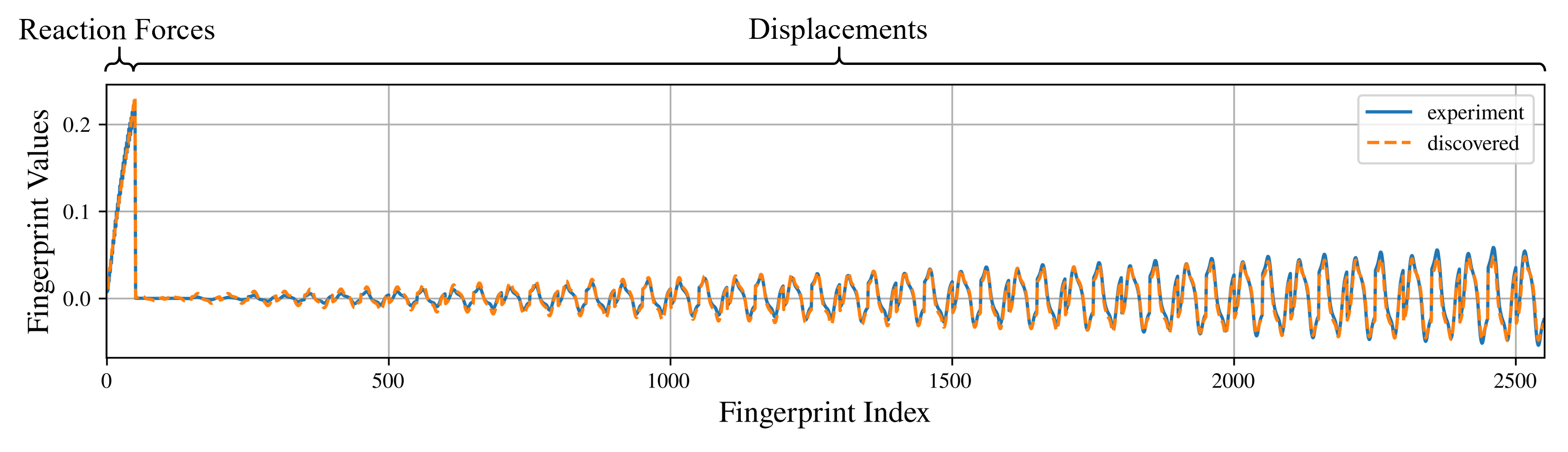}
        \caption{Elastosil 1:1.}
    \end{subfigure}
    \begin{subfigure}[b]{0.7\textwidth}
        \centering
        \includegraphics[width=\textwidth]{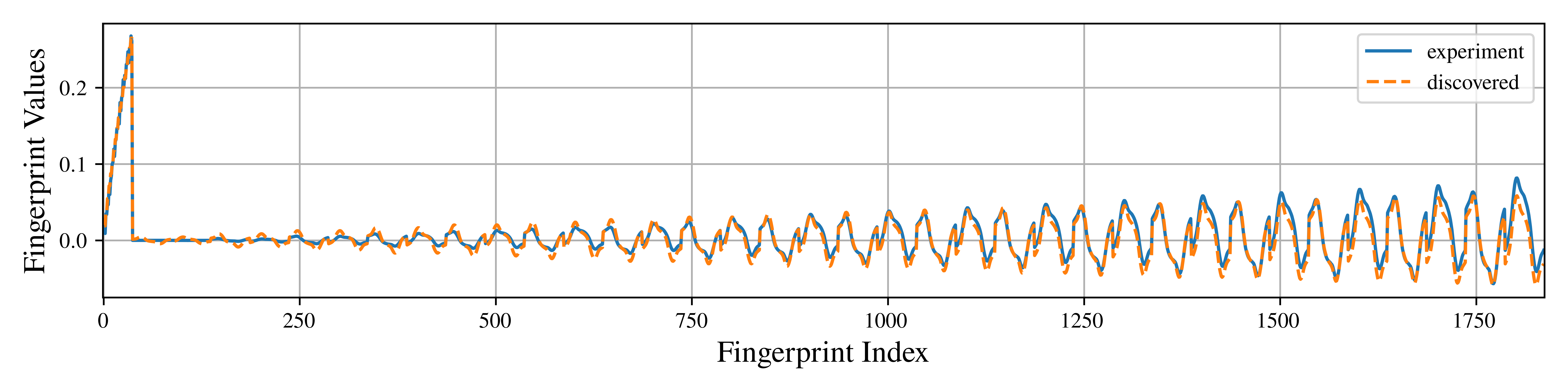}
        \caption{Elastosil 2:1.}
    \end{subfigure}
    \begin{subfigure}[b]{0.7\textwidth}
        \centering
        \includegraphics[width=\textwidth]{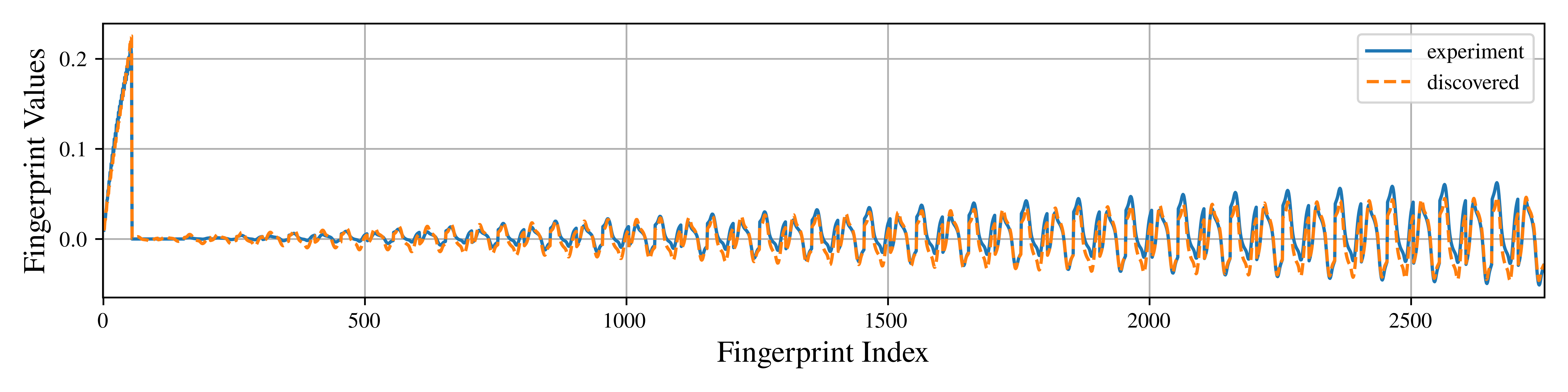}
        \caption{Elastosil 8:5.}
    \end{subfigure}
    \begin{subfigure}[b]{0.7\textwidth}
        \centering
        \includegraphics[width=\textwidth]{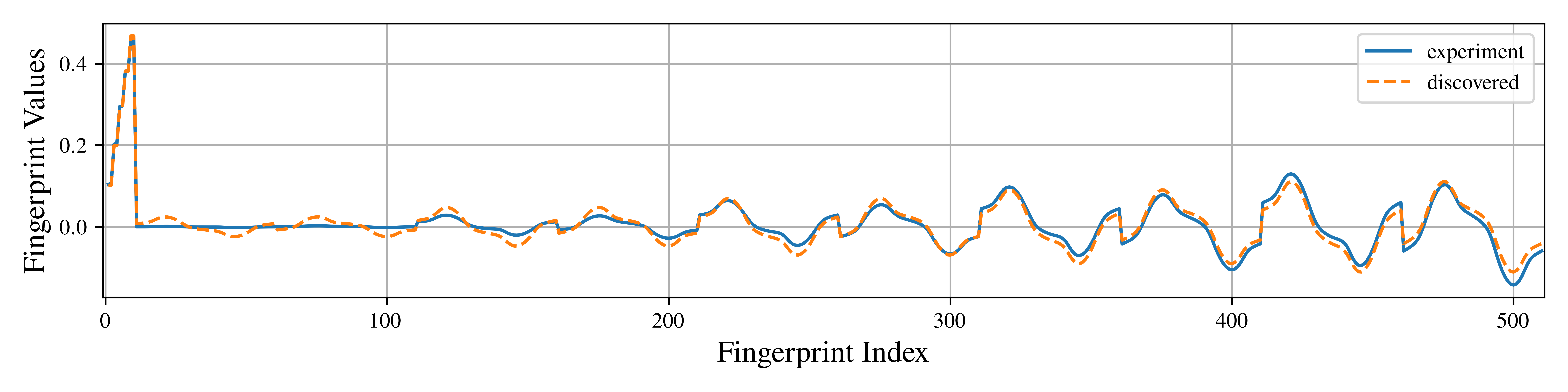}
        \caption{Sylgard.}
    \end{subfigure}
    \begin{subfigure}[b]{0.7\textwidth}
        \centering
        \includegraphics[width=\textwidth]{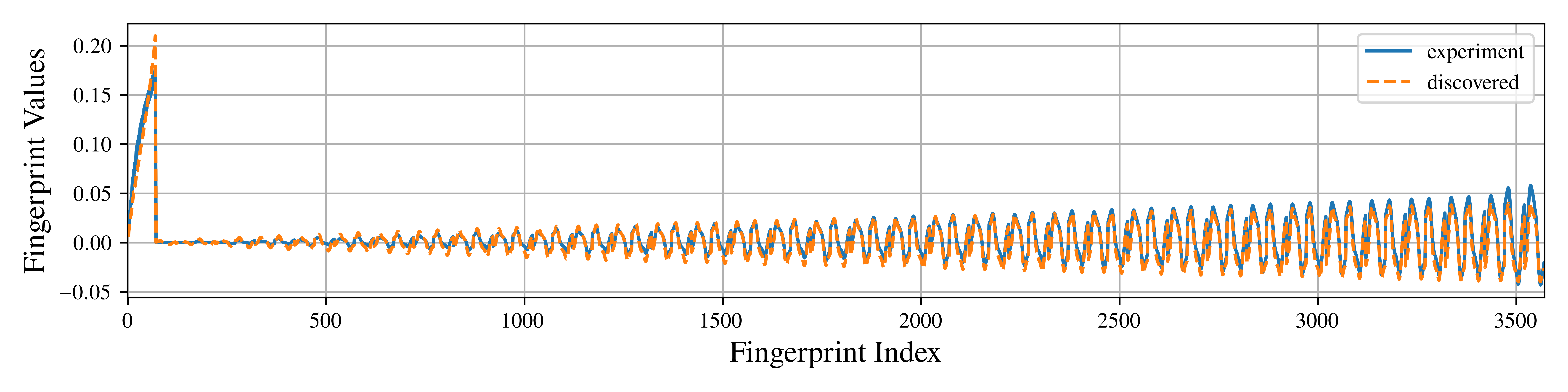}
        \caption{VHB tape.}
    \end{subfigure}
    \caption{Comparison of the measured normalized fingerprints and the best-matching normalized fingerprints in the database for Material Fingerprinting based on Cosine similarity. Continuous curves are shown for visualization purposes only. In practice, the fingerprints consist of discrete vector entries. The fingerprint index has no physical meaning and is used solely to enable plotting and comparison of the experimental and database fingerprints.}
\label{fig:fingerprint_comparison_cosine_similarity}
\end{figure}

To further illustrate the pattern recognition algorithm based on Cosine similarity, we present in \cref{fig:fingerprint_comparison_cosine_similarity} the measured normalized fingerprints together with the best-matching normalized fingerprints from the database. The vectors are shown as graphs, with the vector index on the horizontal axis and the corresponding vector entries on the vertical axis. The left entries correspond to forces, and the right entries correspond to displacements in the fingerprint vectors. We observe a high level of agreement between the measured and identified fingerprints.

\begin{table}[h!]
\caption{
\CHANGE{Identified} strain energy density functions for Material Fingerprinting based on Cosine similarity.
}
\label{tab:discovered_models_cosine_similarity}
\centering
\begin{tabular}{|l|l|c|}
\hline
Materials\vphantom{$\int_{\int}^{\int}$} & Models & Strain Energy Density Functions $\bar{W}$ $\left[\qty{}{\newton \per \milli \meter \squared}\right]$ \\ \hline

\multicolumn{1}{|l|}{Elastosil 1:1\vphantom{$\int_a^b$}} & Mooney-Rivlin & $0.0220 \left[\bar I_1 - 3\right] + 0.0110\left[\bar I_2 - 3\right]$\\ 

\multicolumn{1}{|l|}{Elastosil 2:1\vphantom{$\int_a^b$}} & Mooney-Rivlin & $0.0020 \left[\bar I_1 - 3\right] + 0.0023\left[\bar I_2 - 3\right]$\\ 

\multicolumn{1}{|l|}{Elastosil 8:5\vphantom{$\int_a^b$}} & Mooney-Rivlin & $0.0119 \left[\bar I_1 - 3\right] + 0.0012\left[\bar I_2 - 3\right]$\\ 

\multicolumn{1}{|l|}{Sylgard\vphantom{$\int_a^b$}} & Gen. Neo-Hookean & $0.1850 \left[\left[1+3.1647[\bar I_1 - 3]\right]^{0.5} - 1\right] $\\ 

\multicolumn{1}{|l|}{VHB tape\vphantom{$\int_a^b$}} & Lopez-Pamies & $0.0590 \left[\bar I_1^{0.7164} - 3^{0.7164}\right]$\\ 

\hline
\end{tabular}
\end{table}

The pattern recognition algorithm identifies the Mooney--Rivlin model for the three Elastosil variants, the generalized Neo-Hookean model for Sylgard, and the Lopez-Pamies model for VHB tape as the most suitable models.
Interestingly, we uncover the same material model for all Elastosil variants where only the parameters vary.
The \CHANGE{identified} material models, strain energy density functions, and identified parameters are summarized in \cref{tab:discovered_models_cosine_similarity}.



\begin{figure}[!ht]
    \centering
    \begin{subfigure}[b]{0.19\textwidth}
        \centering
        \includegraphics[width=\textwidth]{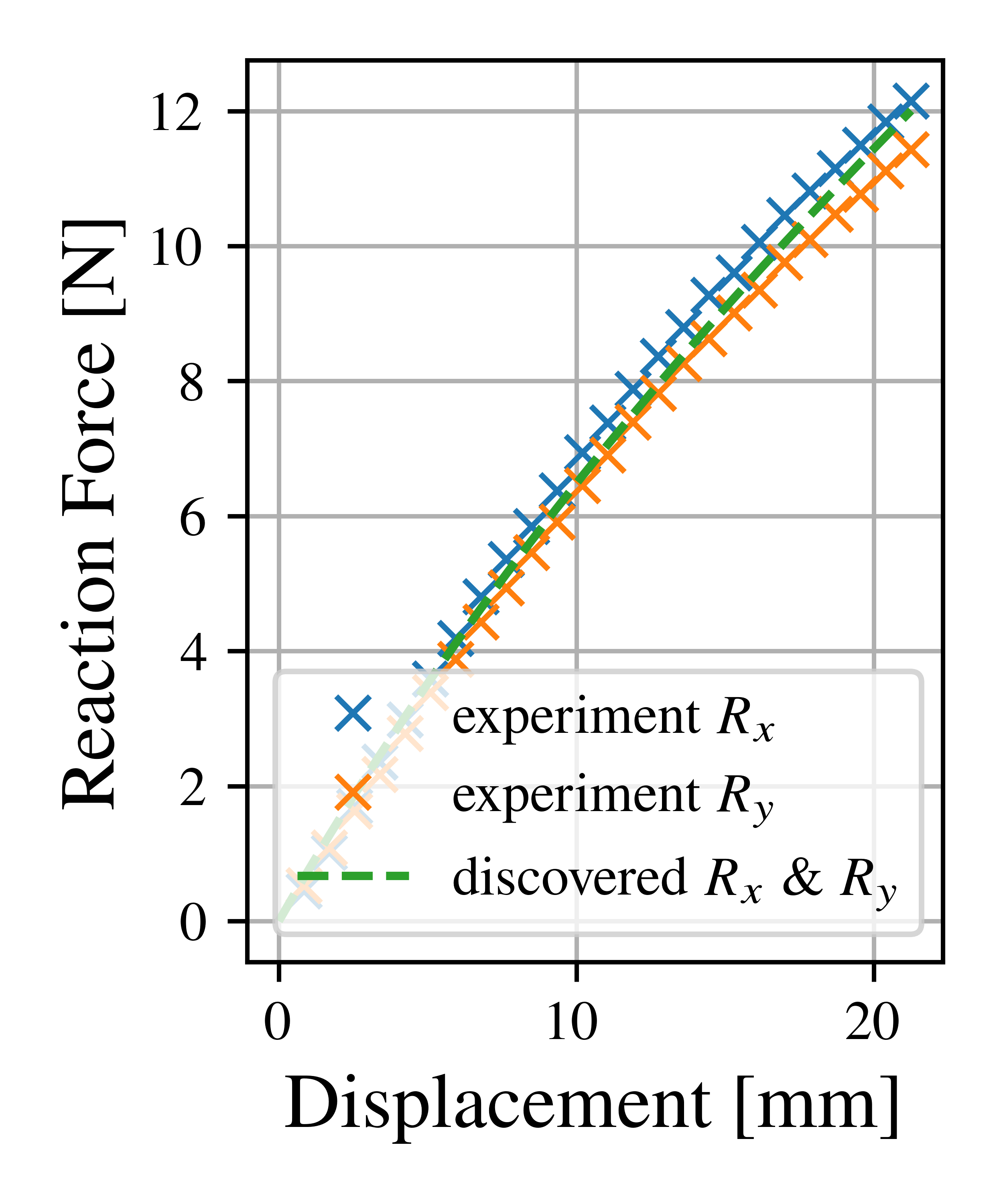}
        \caption{Elastosil 1:1.}
    \end{subfigure}
    \begin{subfigure}[b]{0.19\textwidth}
        \centering
        \includegraphics[width=\textwidth]{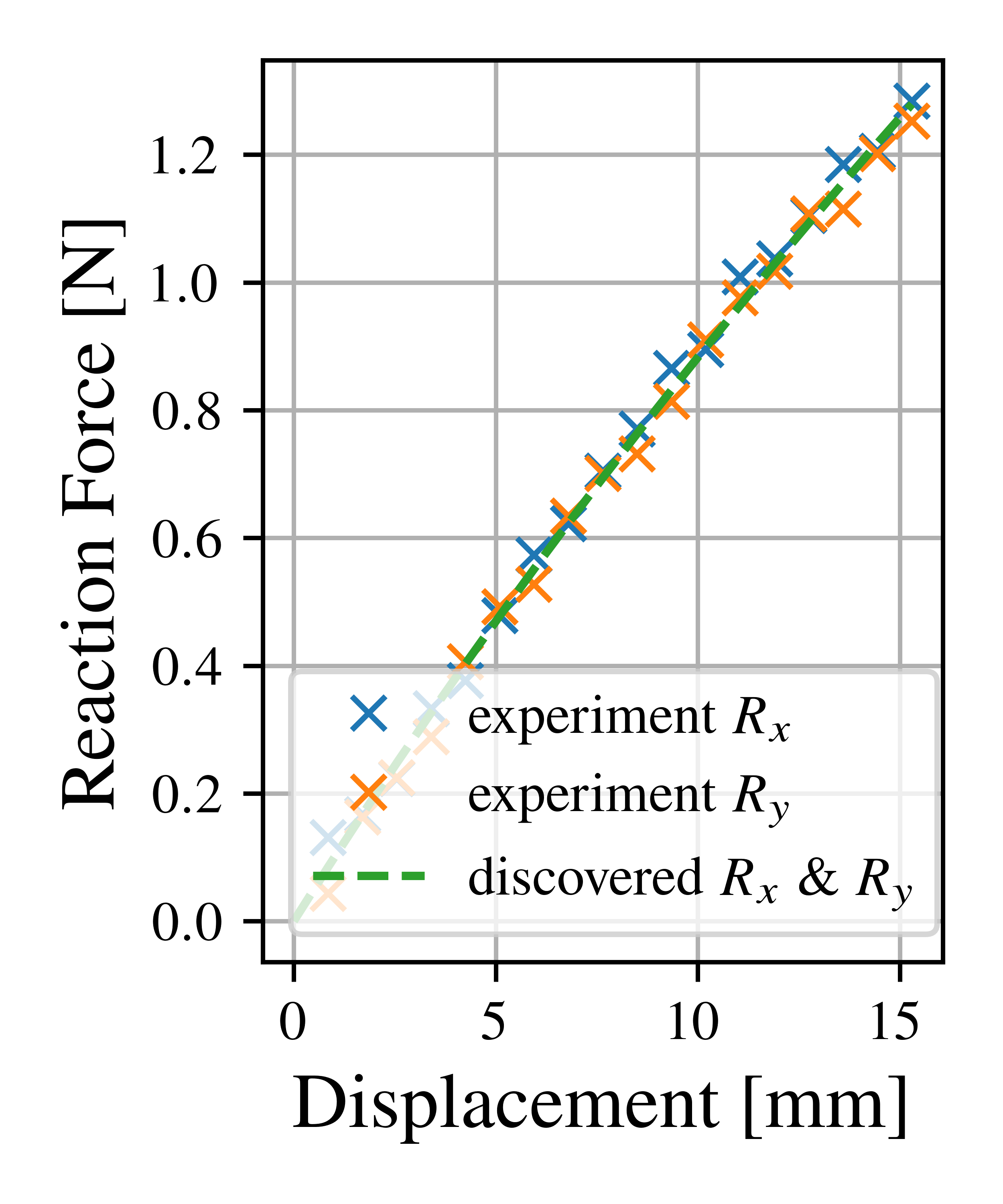}
        \caption{Elastosil 2:1.}
    \end{subfigure}
    \begin{subfigure}[b]{0.19\textwidth}
        \centering
        \includegraphics[width=\textwidth]{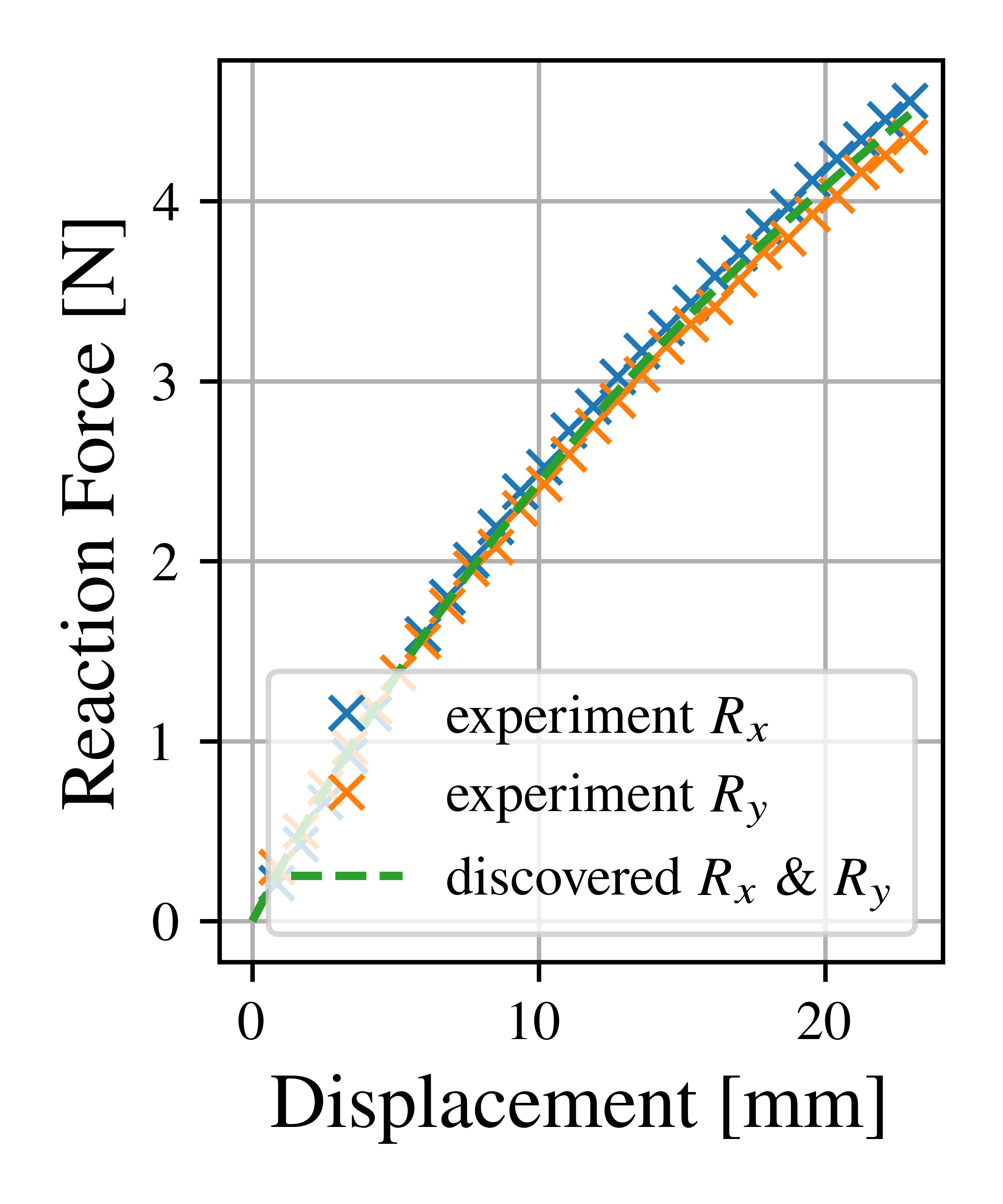}
        \caption{Elastosil 8:5.}
    \end{subfigure}
    \begin{subfigure}[b]{0.19\textwidth}
        \centering
        \includegraphics[width=\textwidth]{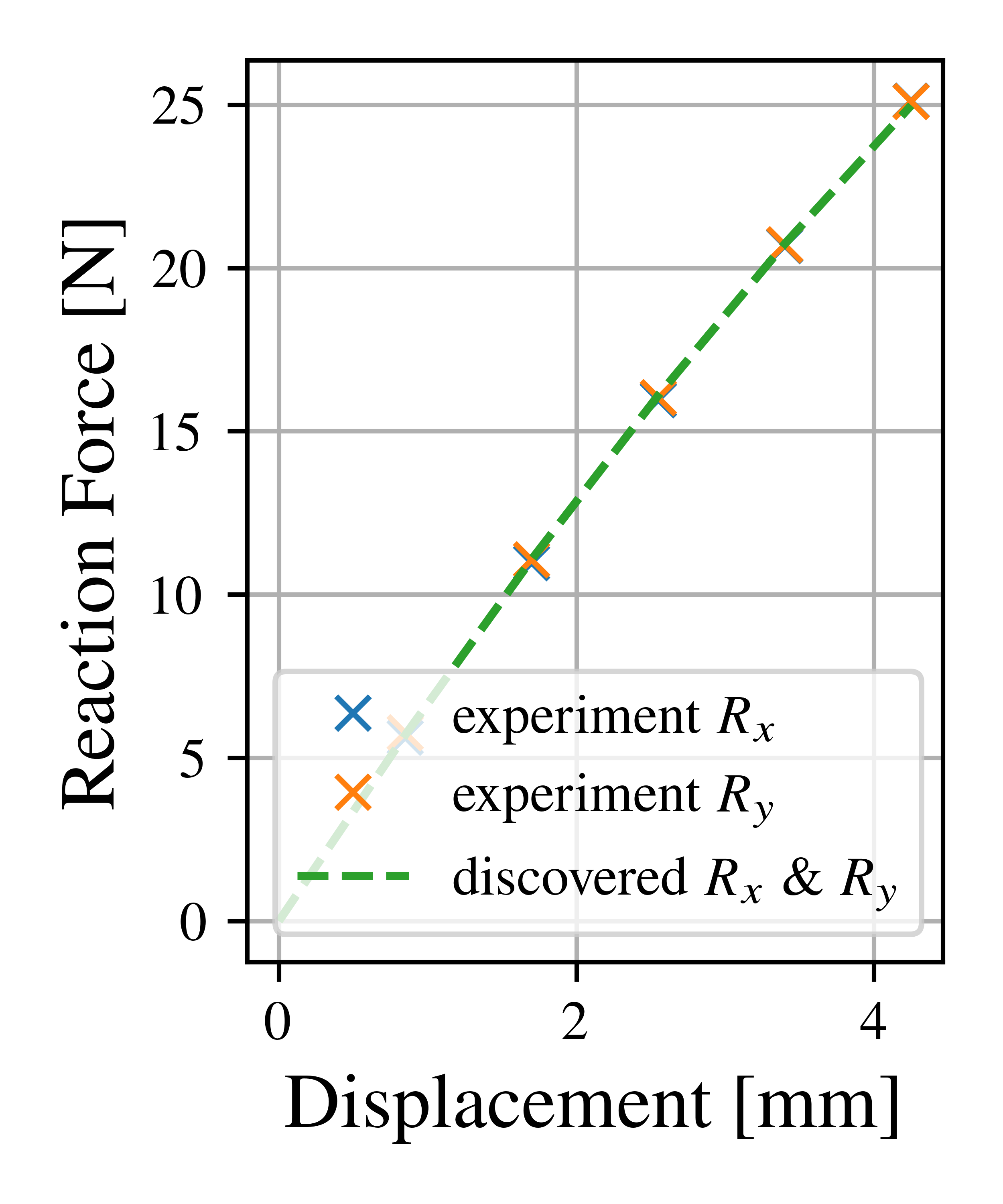}
        \caption{Sylgard.}
    \end{subfigure}
    \begin{subfigure}[b]{0.19\textwidth}
        \centering
        \includegraphics[width=\textwidth]{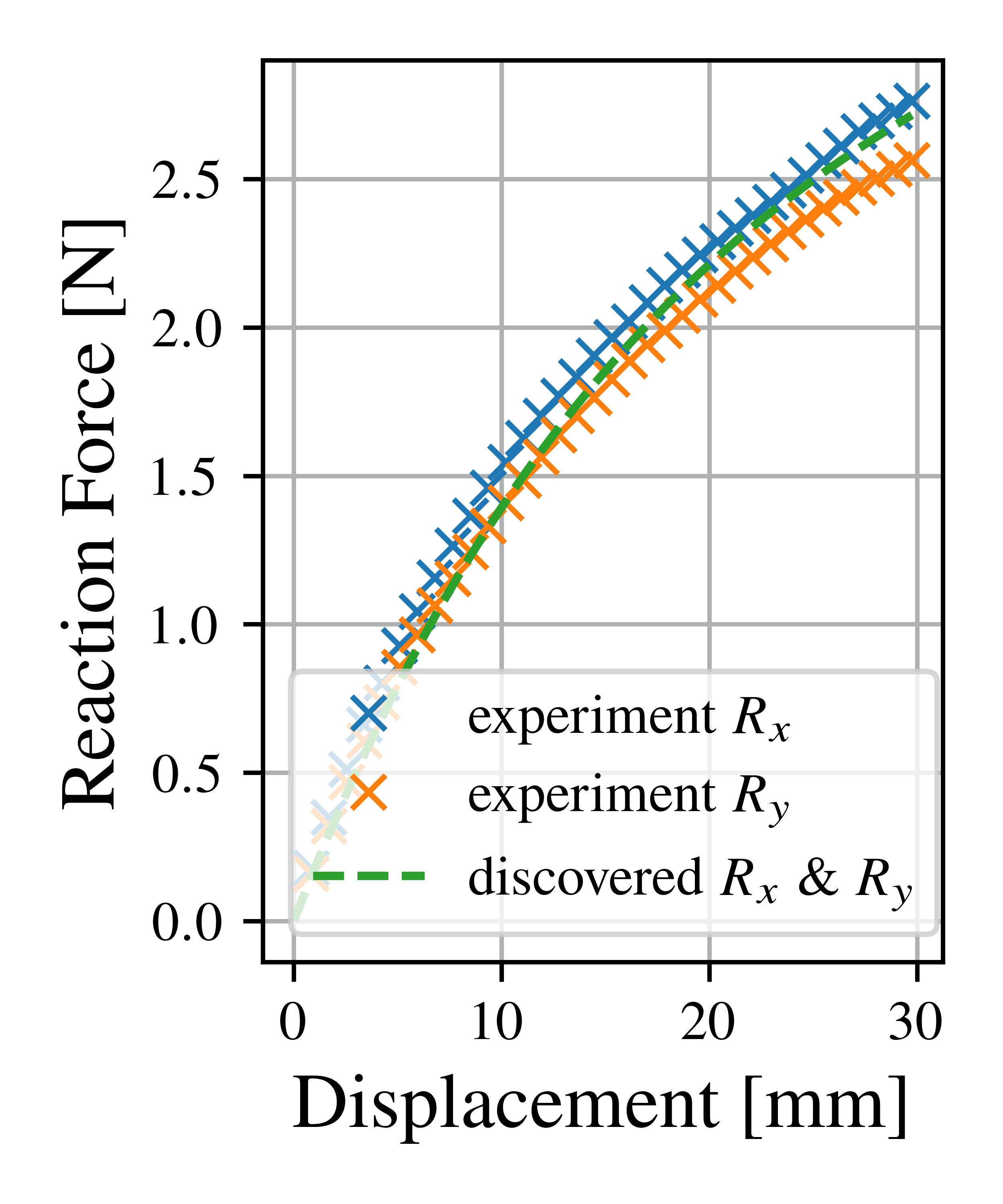}
        \caption{VHB tape.}
    \end{subfigure}
    \caption{Comparison of the measured reaction forces and the reaction forces predicted by the \CHANGE{identified} material models for Material Fingerprinting based on Cosine similarity. The two reaction forces $R_x$ and $R_y$ are obtained from the equibiaxial experiments in \cite{moreno-mateos_biaxial_2025} as the measured forces on each of the two independent axes. Due to experimental variability, the two reaction forces are not exactly identical.}
\label{fig:reaction_forces_cosine_similarity}
\end{figure}

\begin{figure}[ht]
    \centering
    \includegraphics[width=0.9\linewidth]{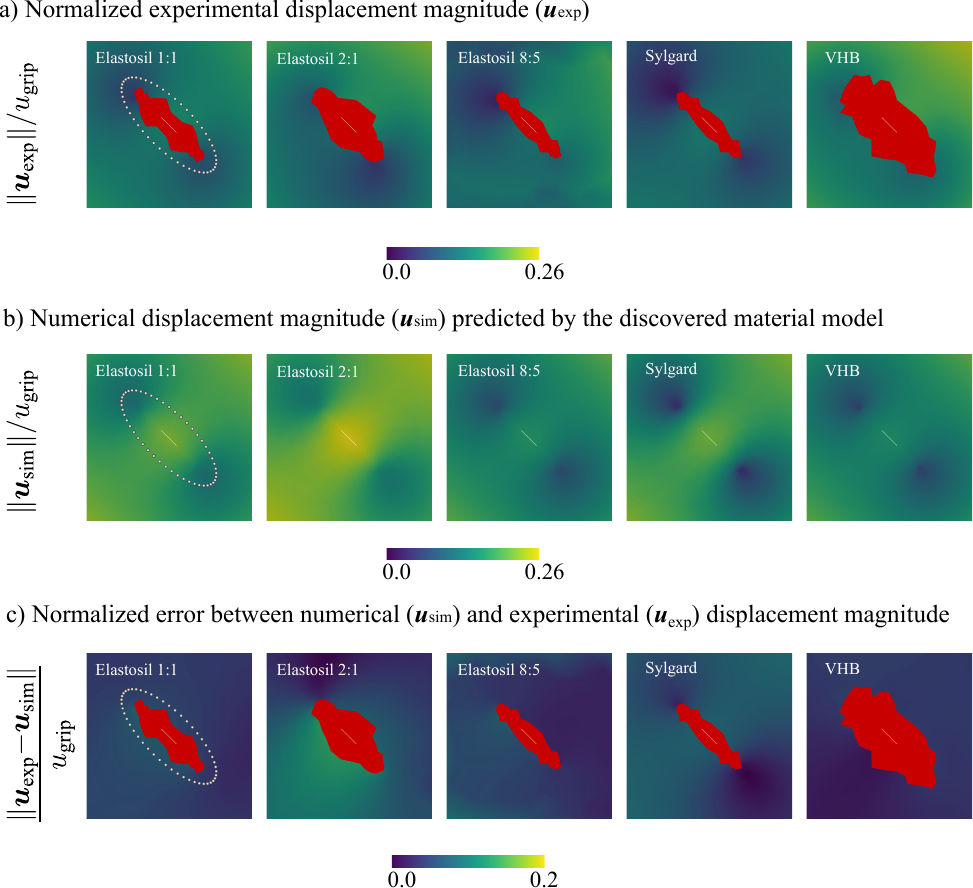}
    \caption{Normalized magnitudes of the experimental (a) and numerical (b) displacement fields in the notch vicinity for Material Fingerprinting based on Cosine similarity, obtained by dividing the displacement magnitude $\lVert \boldsymbol u_\bullet \rVert$ by the grip displacement right before crack onset $(u_{\mathrm{grip}})$: \qty{29.75}{\milli\meter} for VHB tape, \qty{4.25}{\milli\meter} for Sylgard, \qty{15.30}{\milli\meter} for Elastosil 2:1, \qty{22.95}{\milli\meter} for Elastosil 8:5, and \qty{21.25}{\milli\meter} for Elastosil 1:1. Panel~(c) shows the normalized magnitude of the displacement error, computed as $\lVert \boldsymbol u_{\mathrm{sim}} - \boldsymbol u_{\mathrm{exp}} \rVert / u_{\mathrm{grip}}$. The numerical predictions for each material are obtained using the corresponding material model listed in \cref{tab:database_unsupervised}. Experimental displacement fields $\boldsymbol u_{\mathrm{exp}}$ were aggregated across the four experimental repetitions using the geometric mean. The chosen identification nodes are illustrated for Elastosil 1:1 as an example, but are identical for all materials. The red ellipsoidal regions near the center of the contour plots correspond to areas where DIC data is unavailable.
}
\label{fig:displacement_fields_cosine_similarity}
\end{figure}

To assess the fitting accuracy of the model \CHANGE{identified} by \MF{}, we compare the measured reaction forces and displacements to those predicted by the models in \cref{tab:discovered_models_cosine_similarity}.
\cref{fig:reaction_forces_cosine_similarity} shows the experimentally measured and \CHANGE{identified} reaction forces in the horizontal and vertical directions.
For each clamp loading stage, the experimental reaction forces across the four repetitions were aggregated using the geometric mean.
Overall, the reaction forces predicted by the \CHANGE{identified} models agree well with the experimental scatter.
As stated earlier, we observe an almost linear response for Sylgard, which explains why many models exhibit nearly identical Cosine similarity values in this case, see \cref{fig:cosine_similarity}.

\cref{fig:displacement_fields_cosine_similarity} shows the experimentally measured (a) and \CHANGE{identified} (b) displacement fields. Analogous to the reaction forces, the experimental displacement fields for each clamp loading stage were also aggregated using the geometric mean.
While reaction forces are shown for the entire range before crack onset, the displacement fields are presented only for the clamp loading stage at crack onset: \qty{29.75}{\milli\metre} for VHB tape, \qty{4.25}{\milli\metre} for Sylgard, \qty{15.30}{\milli\metre} for Elastosil~2:1, \qty{22.95}{\milli\metre} for Elastosil~8:5, and \qty{21.25}{\milli\metre} for Elastosil~1:1.
The error metric shown in \cref{fig:displacement_fields_cosine_similarity}c corresponds to the absolute error, normalized by the respective clamp loading stages listed above. The contour plots are restricted to a \qty{20}{\milli\metre}\,×\,\qty{20}{\milli\metre} box surrounding the initial notch. Overall, the resulting errors remain small; however, some local disagreement between the experiment and model is observed. This is expected, as the Cosine similarity metric used for model \CHANGE{identification} primarily emphasizes the agreement of displacement directions rather than their absolute magnitude.

\begin{figure}[!ht]
    \centering
    \begin{subfigure}[b]{0.4\textwidth}
        \centering
        \includegraphics[width=\textwidth]{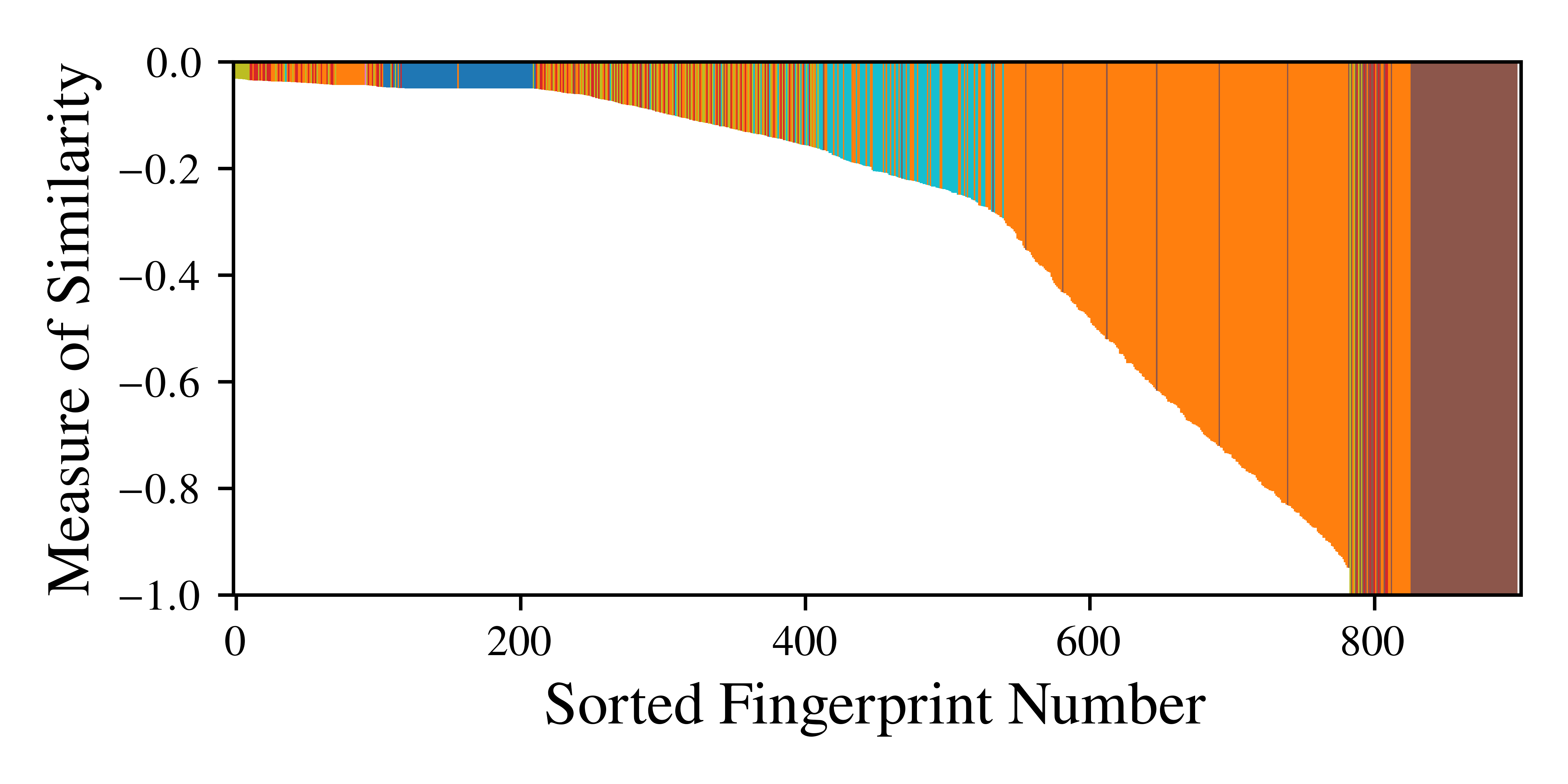}
        \caption{Elastosil 1:1.}
    \end{subfigure}
    \begin{subfigure}[b]{0.4\textwidth}
        \centering
        \includegraphics[width=\textwidth]{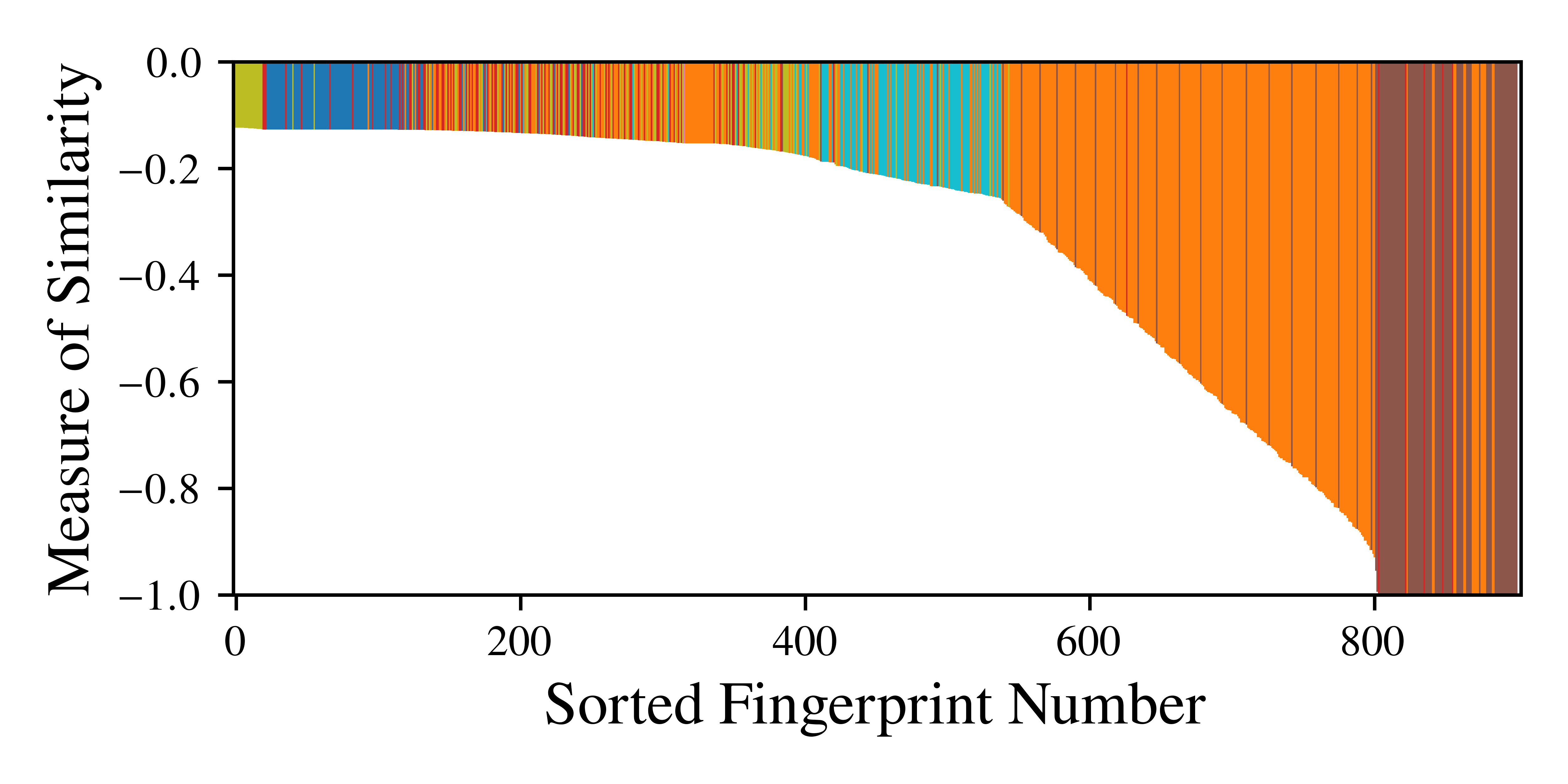}
        \caption{Elastosil 2:1.}
    \end{subfigure}
    \begin{subfigure}[b]{0.4\textwidth}
        \centering
        \includegraphics[width=\textwidth]{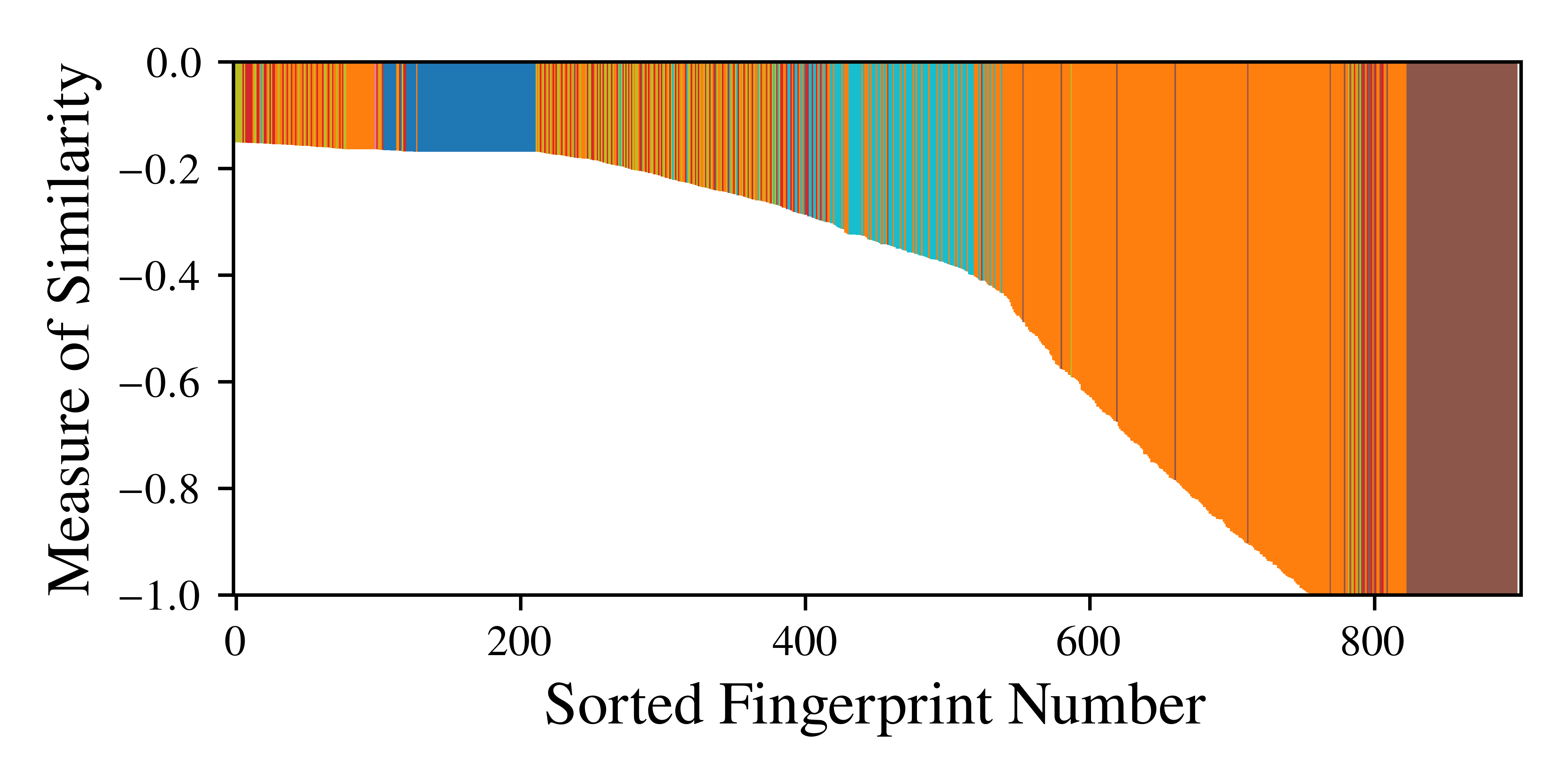}
        \caption{Elastosil 8:5.}
    \end{subfigure}
    \begin{subfigure}[b]{0.4\textwidth}
        \centering
        \includegraphics[width=\textwidth]{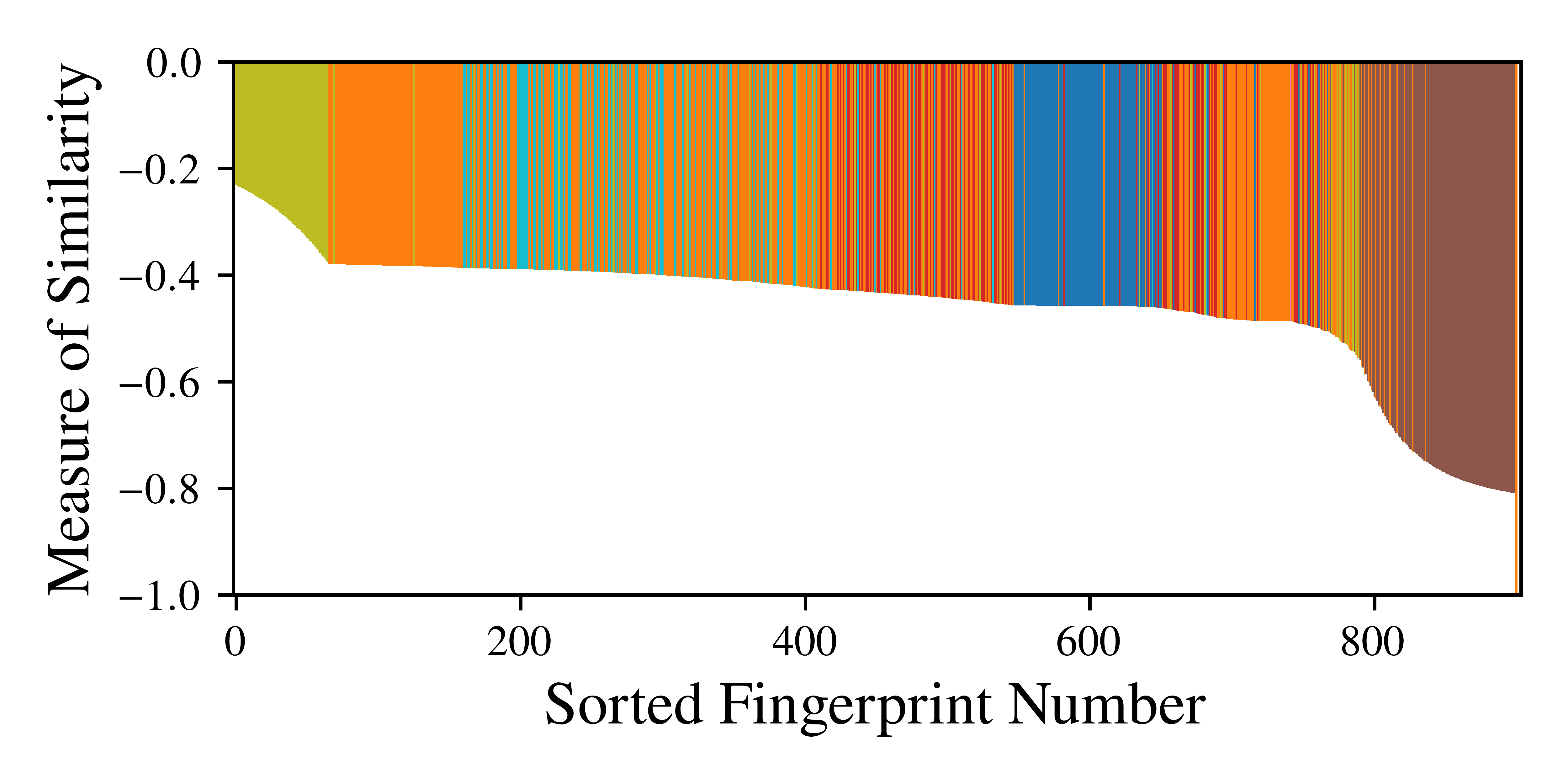}
        \caption{Sylgard.}
    \end{subfigure}
    \begin{subfigure}[b]{0.4\textwidth}
        \centering
        \includegraphics[width=\textwidth]{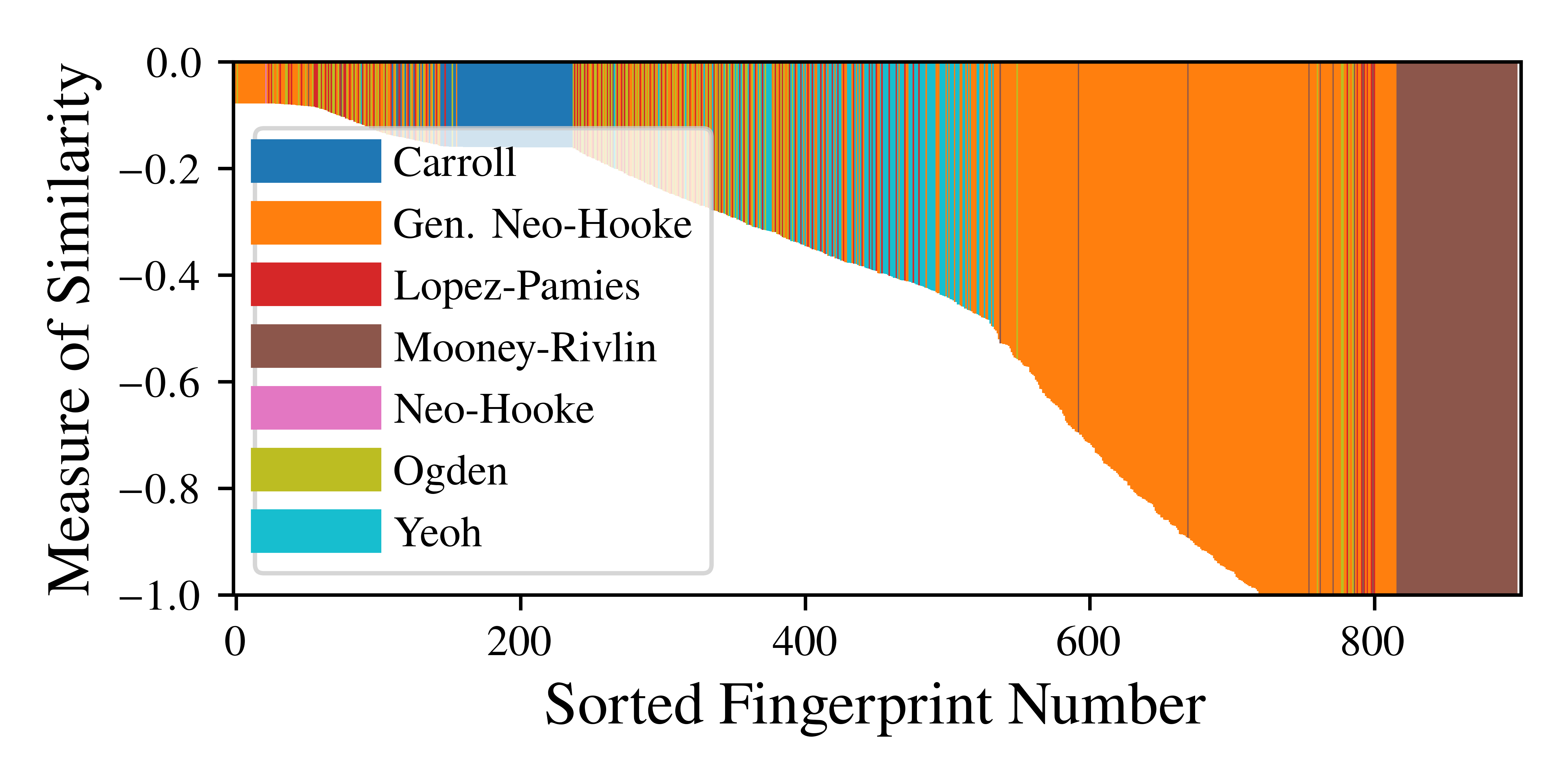}
        \caption{VHB tape.}
    \end{subfigure}
\caption{Illustration of the pattern recognition algorithm for Material Fingerprinting based on Euclidean similarity. The plots displays the similarity measures for all fingerprints, sorted in descending order. The vertical axis is limited to the range $[-1, 0]$ to improve visibility. A value of zero indicates perfect similarity.}
\label{fig:eucidean_similarity}
\end{figure}

\subsection{Euclidean similarity}

As already mentioned, using Cosine similarity for pattern recognition in unsupervised \MF{} does not account for the magnitudes of the displacement fingerprints. Therefore, we introduce the Euclidean similarity in \cref{eq:pattern_recognition_unsupervised_Euclidean} as an alternative similarity measure. In the following, we discuss the results obtained with \MF{} based on Euclidean similarity.

\cref{fig:eucidean_similarity} shows the Euclidean similarities for all fingerprints in the database, ordered by magnitude, with the most suitable fingerprint shown on the left. The use of Euclidean similarity yields a larger variation across all fingerprints. Yet, the top-ranked fingerprints remain considerably ambiguous. Under this measure of similarity, the fingerprints associated with the Mooney-Rivlin model are consistently identified as the worst ones. Interestingly, for Sylgard, there is a convincing match for the Ogden-model-based fingerprints. This contrasts with the inconclusive and ambiguous results obtained using Cosine similarity. Overall, the Euclidean similarity measure is more discriminative, as reflected by the larger variation across fingerprints.

\cref{tab:discovered_models_cosine_similarity} summarizes the \CHANGE{identified} models, their strain energy density functions, and the calibrated material parameters. For the three variants of Elastosil, as well as for Sylgard, the Ogden model is identified. Consistent with the results obtained using Cosine similarity, the same model is \CHANGE{identified} for all Elastosil variants, albeit with different material parameters. For the VHB tape, the generalized Neo-Hookean model is identified.

\begin{table}[h!]
\caption{
\CHANGE{Identified} strain energy density functions for Material Fingerprinting based on Euclidean similarity.
}
\label{tab:discovered_models_cosine_similarity}
\centering
\begin{tabular}{|l|l|c|}
\hline
Materials\vphantom{$\int_{\int}^{\int}$} & Models & Strain Energy Density Functions $\bar{W}$ $\left[\qty{}{\newton \per \milli \meter \squared}\right]$ \\ \hline

\multicolumn{1}{|l|}{Elastosil 1:1\vphantom{$\int_a^b$}} & Ogden & $0.0218 [\bar \lambda_1^{2.6364} + \bar \lambda_2^{2.6364} + \bar \lambda_3^{2.6364} - 3]$\\ 

\multicolumn{1}{|l|}{Elastosil 2:1\vphantom{$\int_a^b$}} & Ogden & $0.0015 [\bar \lambda_1^{3.7273} + \bar \lambda_2^{3.7273} + \bar \lambda_3^{3.7273} - 3]$\\ 

\multicolumn{1}{|l|}{Elastosil 8:5\vphantom{$\int_a^b$}} & Ogden & $0.0074 [\bar \lambda_1^{2.7273} + \bar \lambda_2^{2.7273} + \bar \lambda_3^{2.7273} - 3]$\\ 

\multicolumn{1}{|l|}{Sylgard\vphantom{$\int_a^b$}} & Ogden & $0.0125 [\bar \lambda_1^{10.0} + \bar \lambda_2^{10.0} + \bar \lambda_3^{10.0} - 3]$\\ 

\multicolumn{1}{|l|}{VHB tape\vphantom{$\int_a^b$}} & Gen. Neo-Hookean & $5.9155 \left[\left[1+0.01[\bar I_1 - 3]\right]^{0.5} - 1\right]$\\ 

\hline
\end{tabular}
\end{table}

\begin{figure}[!ht]
    \centering
    \begin{subfigure}[b]{0.19\textwidth}
        \centering
        \includegraphics[width=\textwidth]{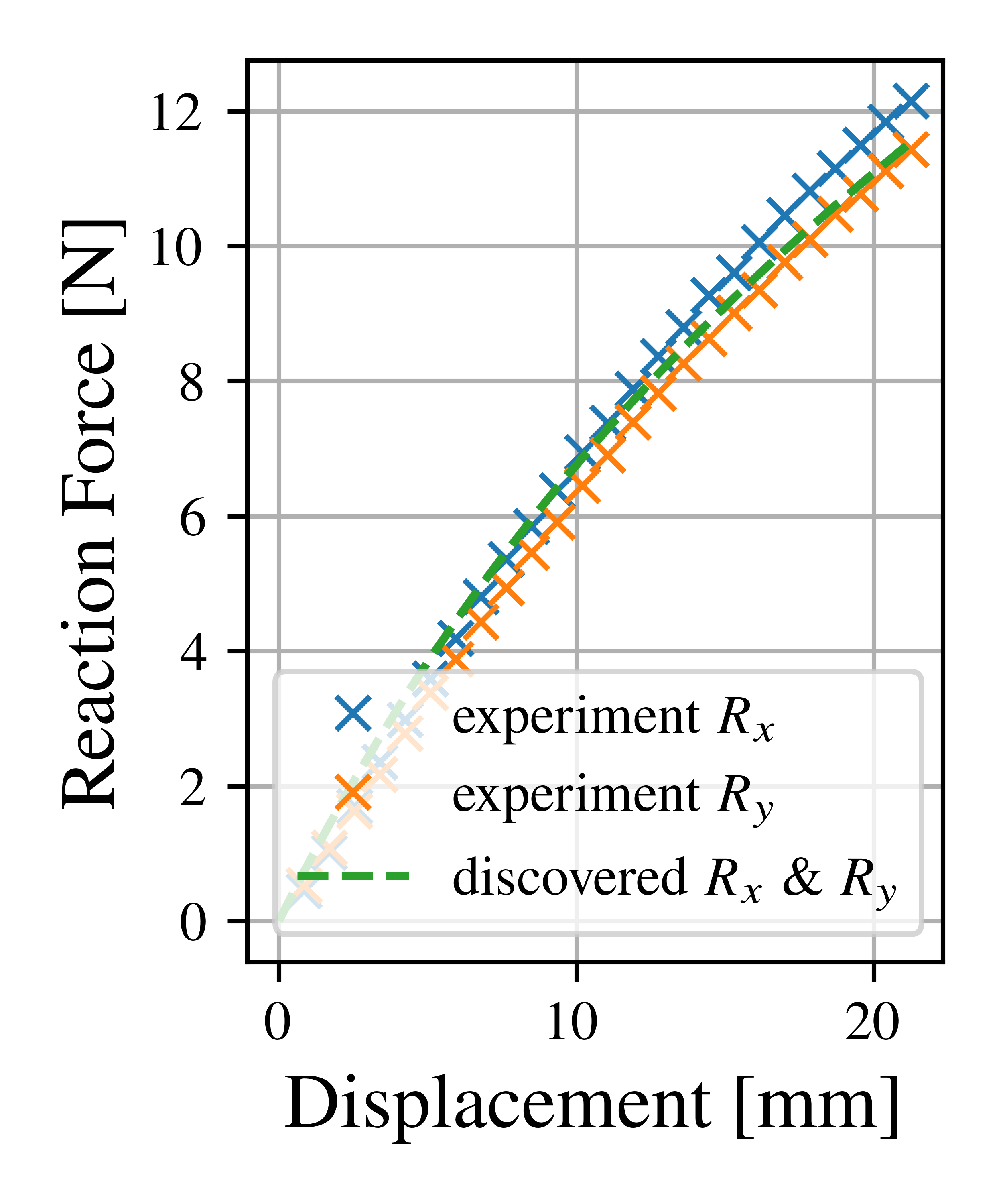}
        \caption{Elastosil 1:1.}
        \label{fig:reaction_forces_euclidean_similarity_elastosil_11}
    \end{subfigure}
    \begin{subfigure}[b]{0.19\textwidth}
        \centering
        \includegraphics[width=\textwidth]{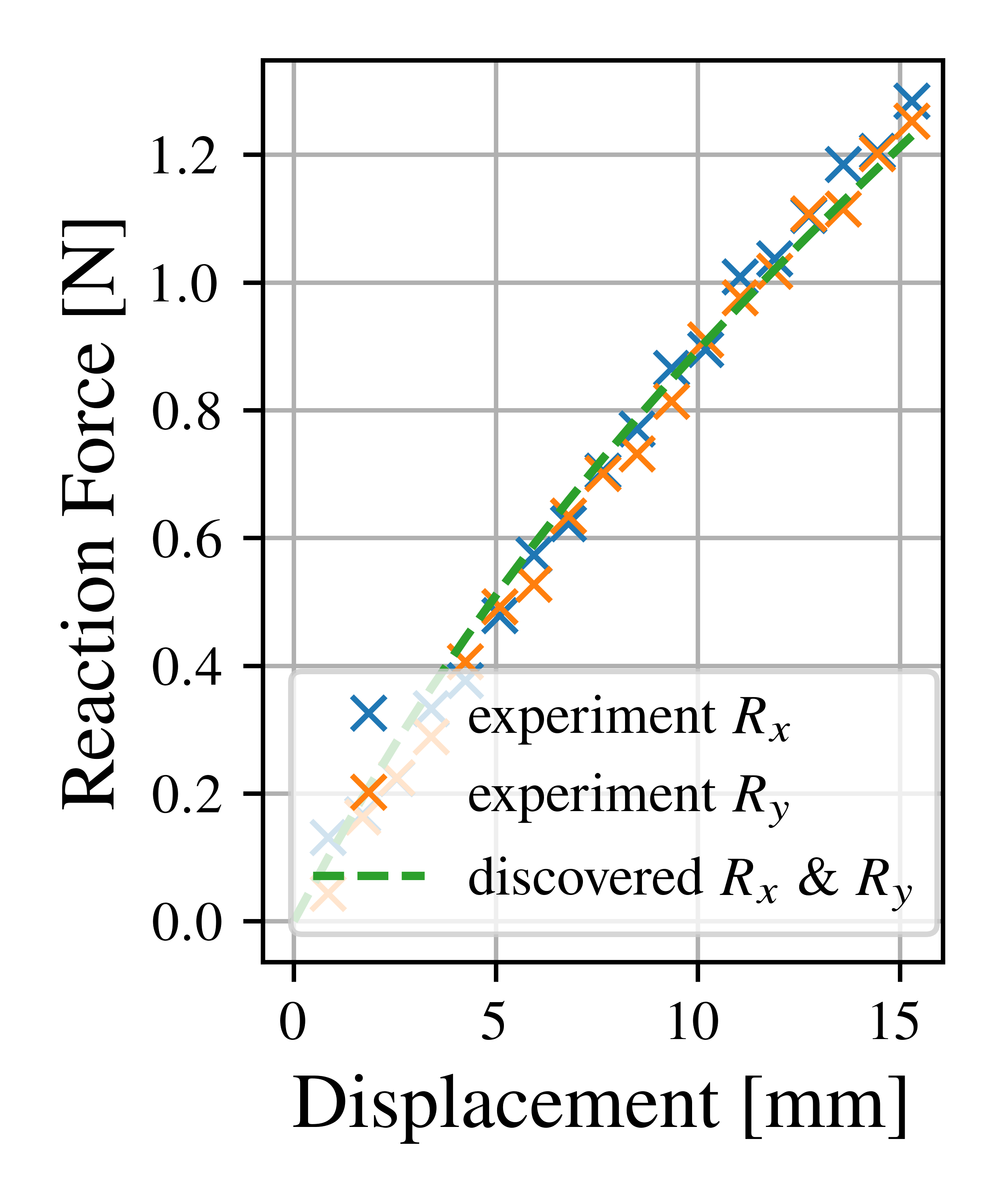}
        \caption{Elastosil 2:1.}
    \end{subfigure}
    \begin{subfigure}[b]{0.19\textwidth}
        \centering
        \includegraphics[width=\textwidth]{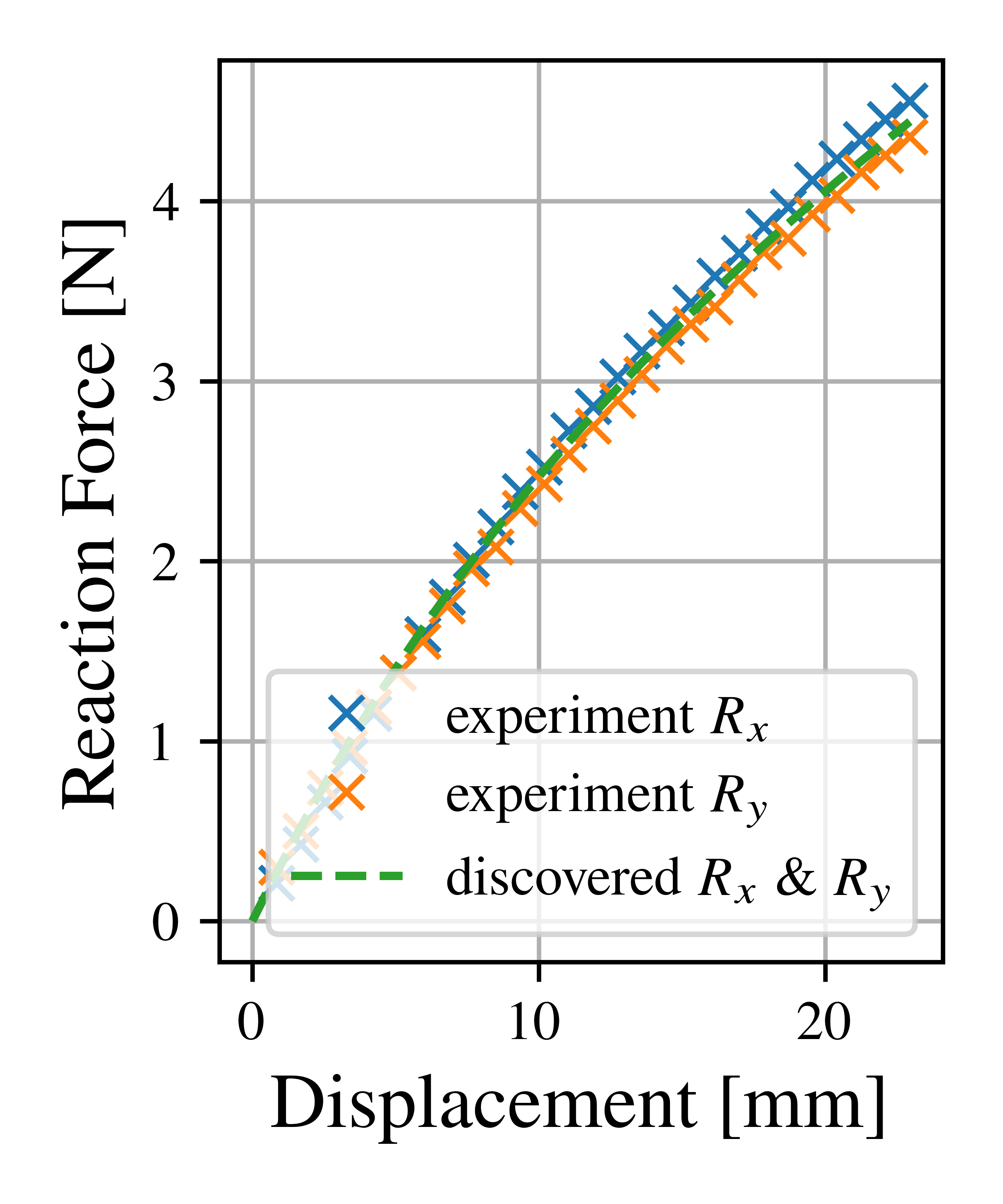}
        \caption{Elastosil 8:5.}
    \end{subfigure}
    \begin{subfigure}[b]{0.19\textwidth}
        \centering
        \includegraphics[width=\textwidth]{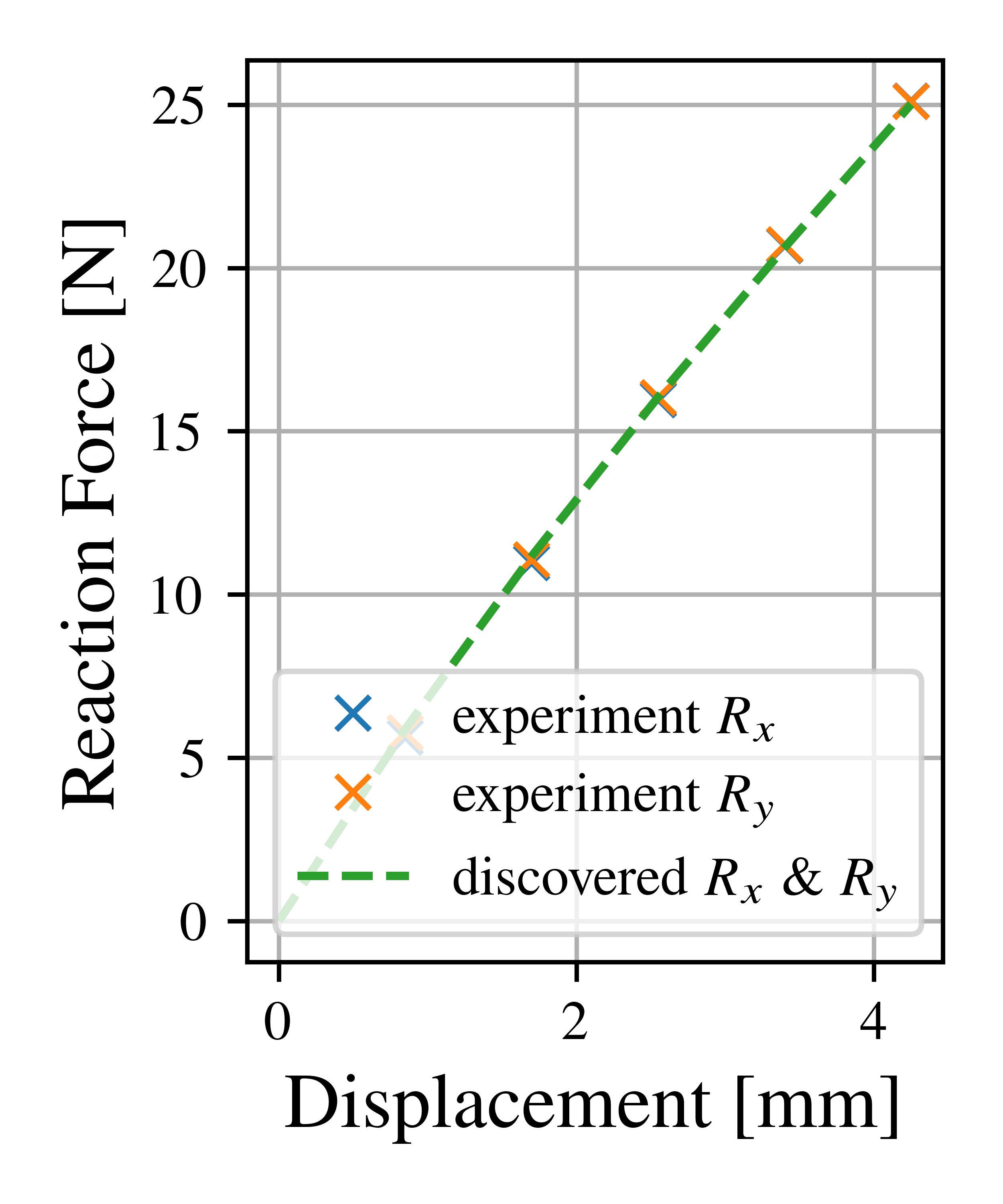}
        \caption{Sylgard.}
    \end{subfigure}
    \begin{subfigure}[b]{0.19\textwidth}
        \centering
        \includegraphics[width=\textwidth]{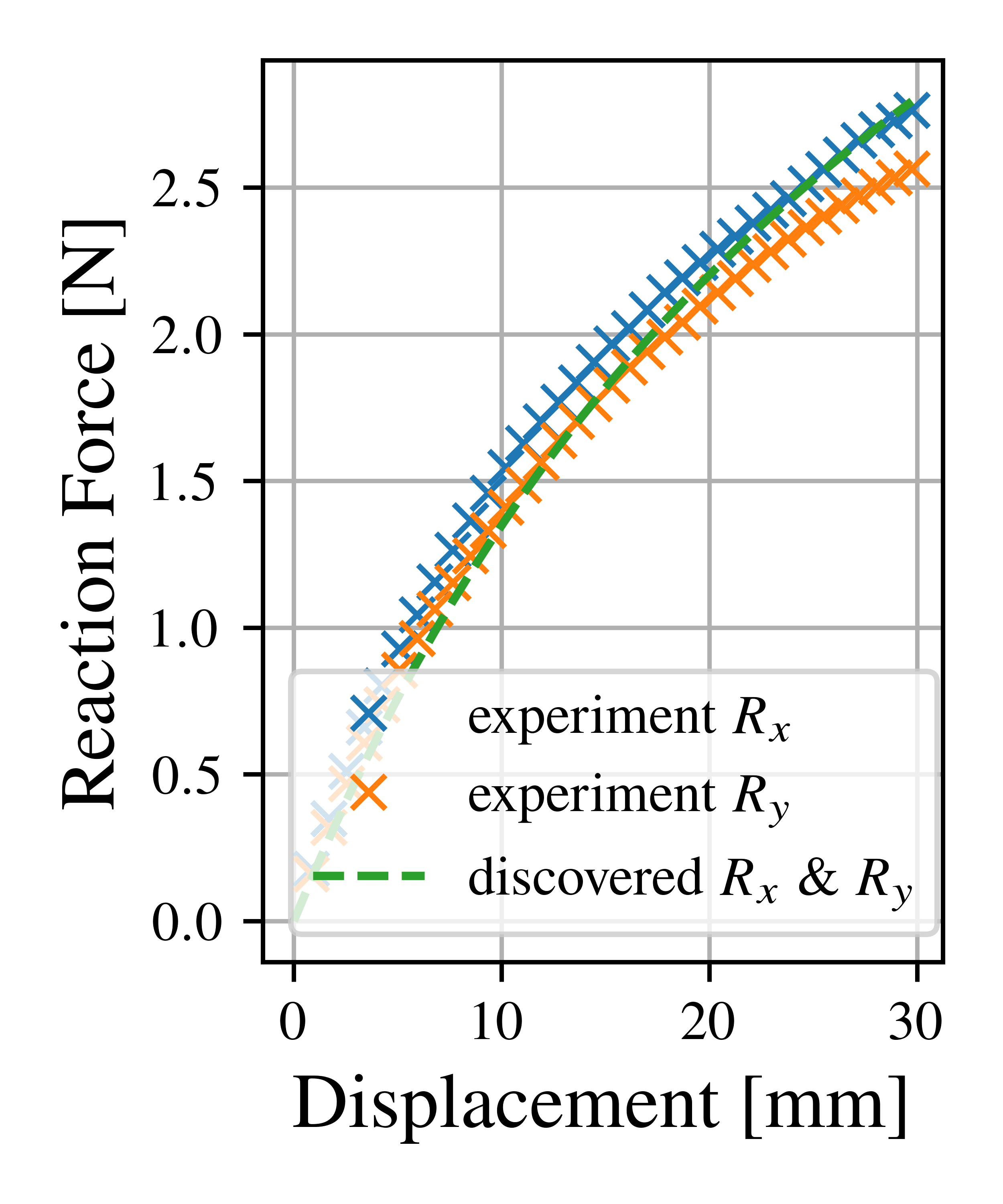}
        \caption{VHB tape.}
    \end{subfigure}
    \caption{Comparison of the measured reaction forces and the reaction forces predicted by the \CHANGE{identified} material models for Material Fingerprinting based on Euclidean similarity. The two reaction forces $R_x$ and $R_y$ are obtained from the equibiaxial experiments in \cite{moreno-mateos_biaxial_2025} as the measured forces on each of the two independent axes. Due to experimental variability, the two reaction forces are not exactly identical.}
\label{fig:reaction_forces_euclidean_similarity}
\end{figure}

To assess the fitting accuracy of the \CHANGE{identified} models, we compare in \cref{fig:reaction_forces_euclidean_similarity} the reaction forces measured in the experiments with those predicted by the identified models. We observe that, similar to the case when Cosine similarity is used, the measured and predicted reaction forces are in good agreement.

\begin{figure}[ht]
    \centering
    \includegraphics[width=0.9\linewidth]{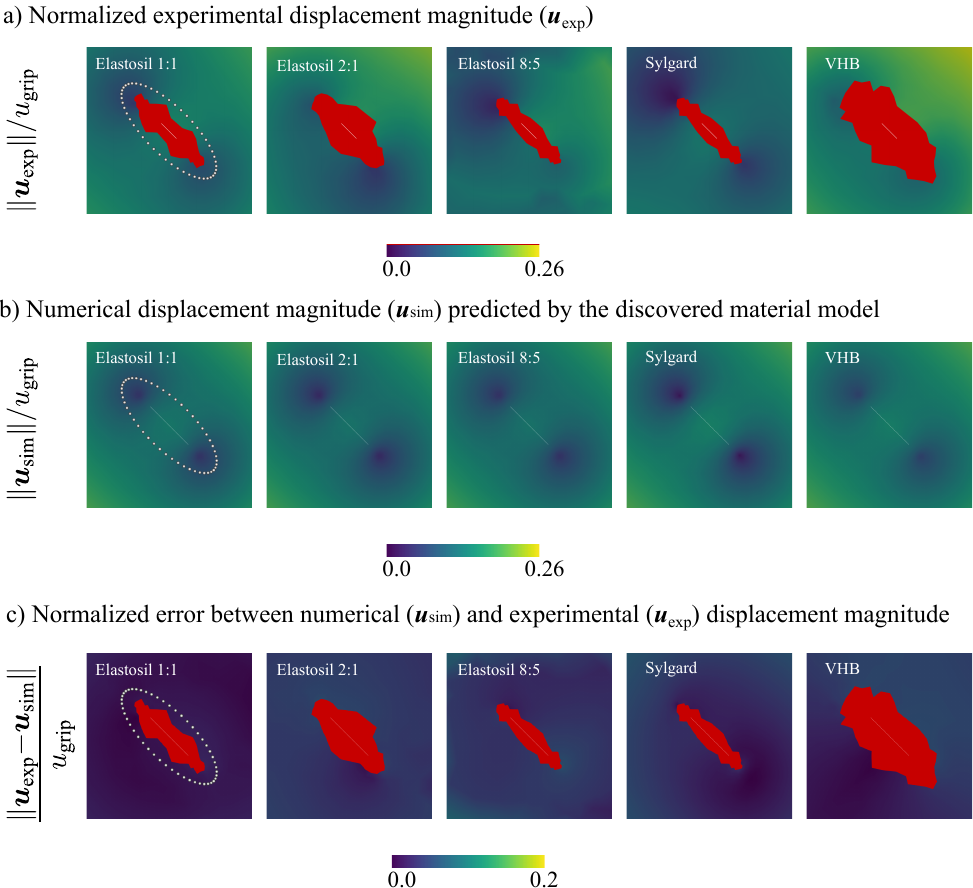}
    \caption{Normalized magnitudes of the experimental (a) and numerical (b) displacement fields in the notch vicinity for Material Fingerprinting based on Euclidean similarity, obtained by dividing the displacement magnitude $\lVert \boldsymbol u_\bullet \rVert$ by the grip displacement right before crack onset $(u_{\mathrm{grip}})$: \qty{29.75}{\milli\meter} for VHB tape, \qty{4.25}{\milli\meter} for Sylgard, \qty{15.30}{\milli\meter} for Elastosil 2:1, \qty{22.95}{\milli\meter} for Elastosil 8:5, and \qty{21.25}{\milli\meter} for Elastosil 1:1. Panel~(c) shows the normalized magnitude of the displacement error, computed as $\lVert \boldsymbol u_{\mathrm{sim}} - \boldsymbol u_{\mathrm{exp}} \rVert / u_{\mathrm{grip}}$. The numerical predictions for each material are obtained using the corresponding material model listed in \cref{tab:database_unsupervised}. Experimental displacement fields $\boldsymbol u_{\mathrm{exp}}$ were aggregated across the four experimental repetitions using the geometric mean. The chosen identification nodes are illustrated for Elastosil 1:1 as an example, but are identical for all materials. The red ellipsoidal regions near the center of the contour plots correspond to areas where DIC data is unavailable.
}
\label{fig:displacement_fields_euclidean_similarity}
\end{figure}

Finally, we compare the experimentally measured displacement fields with those predicted by the identified material models. \cref{fig:displacement_fields_euclidean_similarity} shows the numerically predicted displacement fields (b) in the vicinity of the notch at the loading stage corresponding to crack onset. The experimental displacement contours (a) are consistent with those shown in \cref{fig:displacement_fields_cosine_similarity}. Notably, the normalized error between experimental and predicted displacement fields (c) in \cref{fig:displacement_fields_euclidean_similarity} is significantly smaller than the corresponding errors obtained using Cosine similarity in \cref{fig:displacement_fields_cosine_similarity}c. This effect is particularly pronounced for Elastosil 1:1, for which the errors are almost vanishing. As discussed earlier, Cosine similarity does not account for the magnitudes of displacements. Consequently, the errors are lower when using the Euclidean similarity measure.

\subsection{Computational time and comparison to \CHANGE{continuous optimization methods}}

To assess the computational efficiency of \MF{}, we compare the runtime of its pattern recognition algorithm with that of a traditional optimization-based FEMU procedure. Assuming that the required database has already been generated or is available through an online repository, \MF{} can be performed in under one second on a standard laptop. In contrast, solving an optimization-based FEMU problem, as described in our previous work \citep{moreno-mateos_biaxial_2025}, typically requires approximately five hours. Note that these time estimations are based on computations on an Intel\textsuperscript{\textregistered} Core\texttrademark{} i7-13700 (13th Gen) processor. This comparison indicates that \MF{} is roughly four orders of magnitude faster. It is important to note, however, that these timings depend on the number of material parameters considered in the FEMU formulation and on the spatial discretization of the finite element simulations. Consequently, the reported speedup should be interpreted as a rough estimate of the relative computational effort.

These results demonstrate that, \CHANGE{within the realm of homogeneous, isotropic, and incompressible material behavior considered in this study}, \MF{} offers a computationally efficient alternative to traditional FEMU, achieving inference speeds several orders of magnitude faster once an appropriate database has been constructed. While \MF{} is not intended to replace FEMU, the two approaches are complementary: \MF{} provides rapid predictions, whereas FEMU delivers high-fidelity results through iterative optimization. In practical settings where quick evaluations are essential, such as real-time monitoring, preliminary assessments, or large-scale parametric studies, \MF{} can be especially advantageous. Moreover, \MF{} can serve as an effective strategy for generating high-quality initial guesses that accelerate optimization-based FEMU procedures, \CHANGE{which is investigated in more detail in the next section.} Overall, \MF{} is a strong choice when robustness and computational efficiency are prioritized, whereas FEMU remains preferable when high model accuracy is required.

\CHANGE{Finally, \MF{} should be viewed alongside other full-field identification techniques such as the Virtual Fields Method (VFM) \citep{grediac_principle_1989,pierron_virtual_2012} and the Equilibrium Gap Method (EGM) \citep{claire_finite_2004}, see also the more recent works by \cite{Perotti2017,Makhool2024,Makhool2026}. These methods can also be significantly more efficient than FEMU because they avoid repeated finite element analyses during parameter identification. However, VFM and EGM are often more sensitive to experimental noise. Yet, they provide attractive alternatives when full-field measurements are available and have even been extended to enable constitutive model discovery \citep{abbasi_discovery_2026}. Therefore, the choice among \MF{}, FEMU, VFM, and EGM (or any other method) depends on the trade-off between computational efficiency, robustness to noise, model complexity, and the desired physical fidelity.}

\subsection{\CHANGE{\MF{} as an initializaton strategy for FEMU}}

As discussed above, \MF{} can also be used to provide high-quality initial guesses for optimization-based FEMU. To investigate this possibility, \CHANGETWO{we consider the Elastosil~1:1 dataset and the Ogden model, which was identified as the best-performing constitutive model by \MF{} for the Euclidean similarity (see~\cref{tab:discovered_models_cosine_similarity}). A FEMU calibration is then performed using the \texttt{trust-constr} algorithm from SciPy with the same objective function employed by the Euclidean similarity measure, i.e.,~\cref{eq:pattern_recognition_unsupervised_Euclidean}.} Tight convergence tolerances of $\num{1e-7}$ are prescribed for both \texttt{gtol} and \texttt{xtol}.\footnote{\CHANGE{The tolerances were intentionally chosen to be tight to ensure a fair comparison between both initialization strategies. In practical applications, looser tolerances may often be sufficient, in which case the computational advantage of initializing FEMU with the \MF{} estimate may be even more pronounced.}}

Two initial guesses are considered. The first corresponds to a generic initial guess, $(\theta_8, \alpha_4) = (1.0,2.0)$, while the second is the \MF{} estimate $(\theta_8, \alpha_4) = (0.0218,2.6364)$ reported in~\cref{tab:discovered_models_cosine_similarity}. Both optimizations converge successfully to nearly identical parameter values, namely $(0.02103,2.68095)$ and $(0.02103,2.68111)$, respectively, yielding nearly identical objective function values. \CHANGETWO{The \MF{} estimate is already located very close to the optimum identified by FEMU, with relative parameter changes of approximately 3.3\% in $\theta_8$ and 1.7\% in $\alpha_4$.}

\begin{figure}[t]
    \centering
    \begin{subfigure}[t]{0.48\textwidth}
        \centering
        \includegraphics[width=\textwidth]{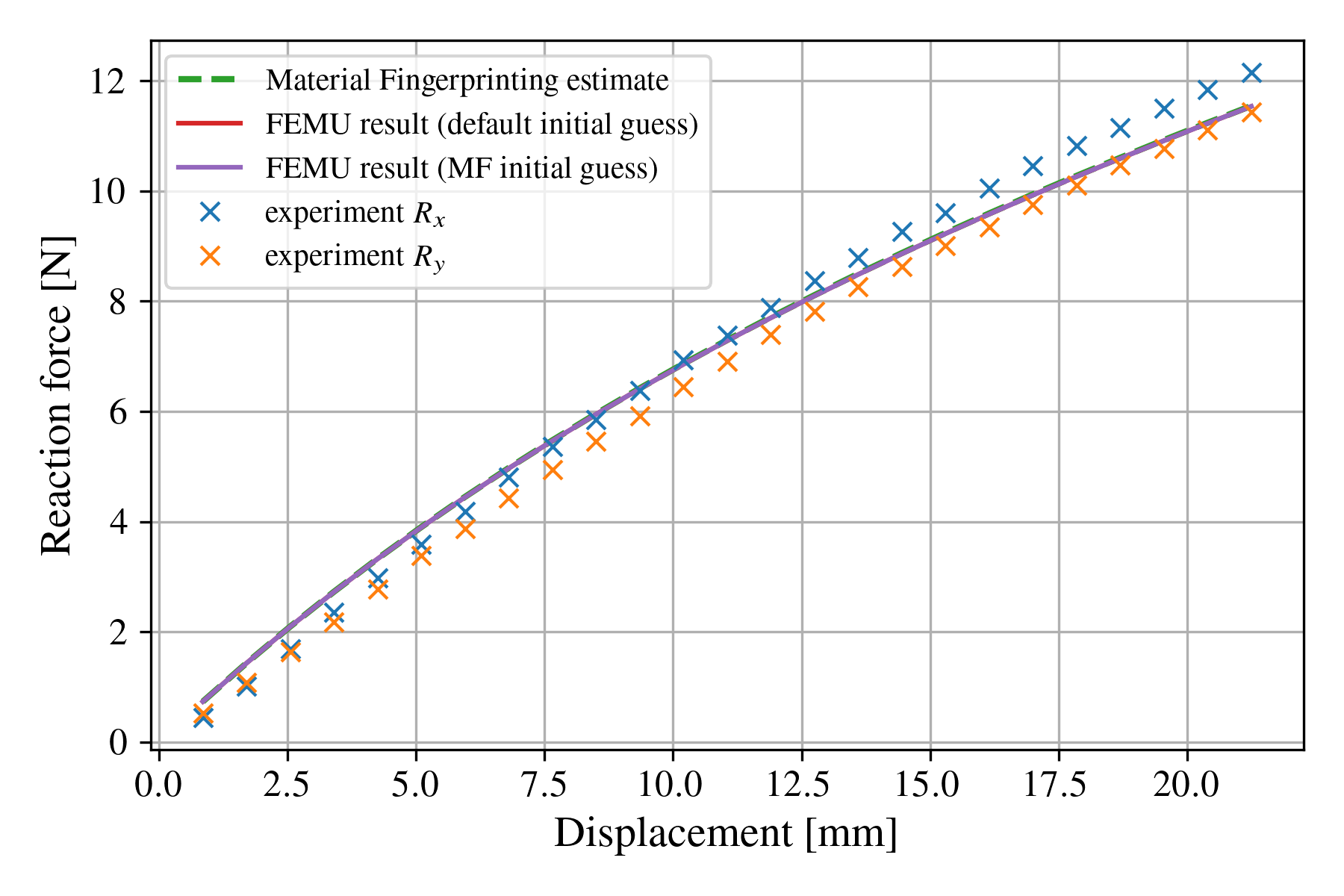}
        \caption{Reaction force comparison.}
        \label{fig:ogden_force_comparison}
    \end{subfigure}
    \hfill
    \begin{subfigure}[t]{0.46\textwidth}
        \centering
        \includegraphics[width=\textwidth]{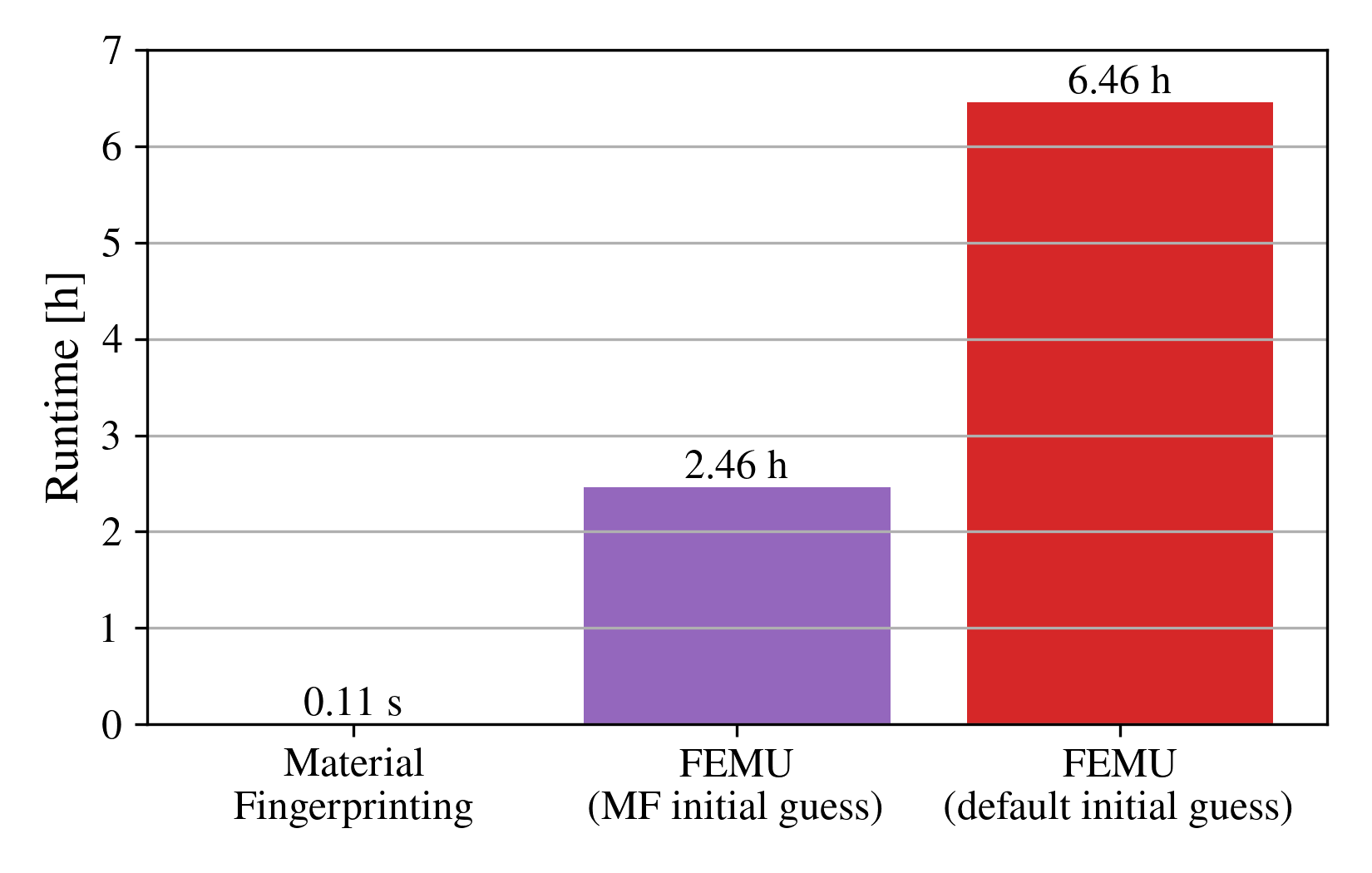}
        \caption{Computational cost comparison.}
        \label{fig:ogden_runtime_comparison}
    \end{subfigure}
    \caption{\CHANGE{(a) Comparison of experimental and reaction forces predicted by the \MF{} estimate from~\cref{tab:discovered_models_cosine_similarity} (Elastosil 1:1, Ogden) and FEMU results using a default initial guess and the \MF{} estimate. (b) Corresponding wall times.}}
    \label{fig:femu_study}
\end{figure}

\CHANGE{However, substantial differences are observed in the computational effort. Starting from the generic initial guess requires 84 objective function evaluations and a wall time of approximately 6.46~h, whereas initialization with the \MF{} estimate requires only 33 objective function evaluations and 2.46~h wall time. This corresponds to a reduction of approximately 61\% in wall time (see~\cref{fig:ogden_runtime_comparison}).}

\CHANGE{\cref{fig:ogden_force_comparison} compares the reaction forces obtained from the two FEMU solutions together with the direct \MF{} estimate, recall~\cref{fig:reaction_forces_euclidean_similarity_elastosil_11}. Although the subsequent FEMU calibration further reduces the objective function, the force-displacement curves obtained from the \MF{} estimate are already in almost perfect agreement with the optimized FEMU solutions. These results indicate that \MF{} not only provides ultra-fast parameter estimates, but can also serve as an effective preprocessing step that significantly accelerates optimization-based parameter identification.}

\CHANGETWO{We finally note that, in the current investigation, the FEMU approach is restricted to the Ogden material model. In contrast, Material Fingerprinting can naturally distinguish between different material models. To achieve the same capability with FEMU, a separate optimization must be performed for each candidate material model.}

\section{Conclusions and outlook}
\label{sec:conclusions}

In this work, we presented the first experimental \CHANGE{demonstration} of unsupervised Material Fingerprinting for the \CHANGE{identification} of hyperelastic constitutive models from real full-field measurements and global reaction forces. By leveraging a precomputed database of simulated fingerprints and a pattern recognition framework, \MF{} enables ultra-fast and robust material model \CHANGE{inference} without solving a \CHANGE{continuous} optimization problem. The proposed approach bypasses common challenges associated with optimization-based identification techniques, including high computational costs and sensitivity to local minima.

We demonstrated the effectiveness of the method using standardized equibiaxial experiments with heterogeneous deformation fields induced by a central notch. A single database of incompressible hyperelastic material models was successfully used to characterize five distinct soft materials based solely on experimentally measured displacements and reaction forces. The identified constitutive models accurately reproduce both global force responses and local displacement fields up to crack initiation. Moreover, we showed that, once the database is available, \MF{} achieves inference times on the order of seconds, representing a speedup of several orders of magnitude compared to traditional FEMU approaches.

Beyond \CHANGE{demonstrating the applicability of} unsupervised \MF{} on experimental data, we introduced and assessed two similarity measures for fingerprint matching. While Cosine similarity emphasizes the shape of the fingerprints, a Euclidean-distance-based metric preserves information about both shape and magnitude and is particularly well suited for fingerprints that combine heterogeneous physical quantities. Both approaches yielded consistent and physically interpretable results, which highlight the robustness and flexibility of the Material Fingerprinting framework. Notably, the Euclidean-based similarity measure provides greater discriminative power, as reflected by the larger variation across fingerprints.
\CHANGETWO{The Euclidean-based similarity measure is well aligned with the FEMU paradigm. However, in conventional FEMU approaches, the computational effort is invested independently for each new material identification task. The large number of finite element simulations required for parameter identification is typically not reused after the constitutive model has been identified. Material Fingerprinting instead treats these simulations as a long-term resource that can be reused for future identification tasks.}


\CHANGETWO{A remaining challenge for future research is the design of standardized, easy-to-follow, and concise testing protocols that enable direct comparison with the fingerprints stored in the database. In this context, one may even envision standardized, openly available, 3D-printable CAD files for fabricating either the test specimens themselves or the molds used to produce them.}

The present study focuses on isotropic, incompressible hyperelastic materials under quasi-static loading. Future work will aim to expand the database to include more complex material behaviors, such as anisotropy \citep{flaschel_adaptive_2026}, compressibility, and rate-dependent or dissipative effects. \CHANGE{We anticipate that extending Material Fingerprinting to dissipative materials will introduce additional challenges, particularly with regard to the efficient sampling of high-dimensional parameter spaces and the resulting increase in database size. To address these challenges, future work will focus on the development and investigation of both linear and nonlinear database compression techniques.} Further, alternative experiments for unsupervised material fingerprinting may include compression-based setups, such as the indentation of soft materials (cf. e.g., the work by \cite{Ashkenazi2025} or the indentation driven by a cutting tool by \cite{MorenoMateos2025-cutting}). \CHANGE{We also note that the \MF{} framework discussed in this work is restricted to discovering models that are present in the fingerprint database, but cannot discover combinations of these models. An extension to an adaptive \MF{} framework that discovers linear combinations of models in the database, as shown by \cite{flaschel_adaptive_2026} in the supervised setting, would be a promising direction for future research.} In addition, incorporating uncertainty quantification, adaptive database refinement, and hybrid strategies that combine \MF{} with optimization-based methods represent promising directions.

\section*{Code and data availability}

The code and data are publicly available on GitHub: 

\url{https://github.com/orgs/Material-Fingerprinting/material-fingerprinting-hyperelasticity-unsupervised-experiments}

We acknowledge the use of the \texttt{dolfiny} implementation of principal stretch computations and their derivatives:

\url{https://github.com/fenics-dolfiny/dolfiny}

\section*{Acknowledgments}
The authors acknowledge support from the European Research Council (ERC) under the Horizon Europe Program, Grant-No. 101141626 DISCOVER and Grant-No. 101052785 SoftFrac. Funded by the European Union. Views and opinions expressed are, however, those of the authors only and do not necessarily reflect those of the European Union or the European Research Council
Executive Agency. Neither the European Union nor the granting authority can be held responsible for them.
\begin{figure}[H]
\includegraphics[width=0.3\textwidth]{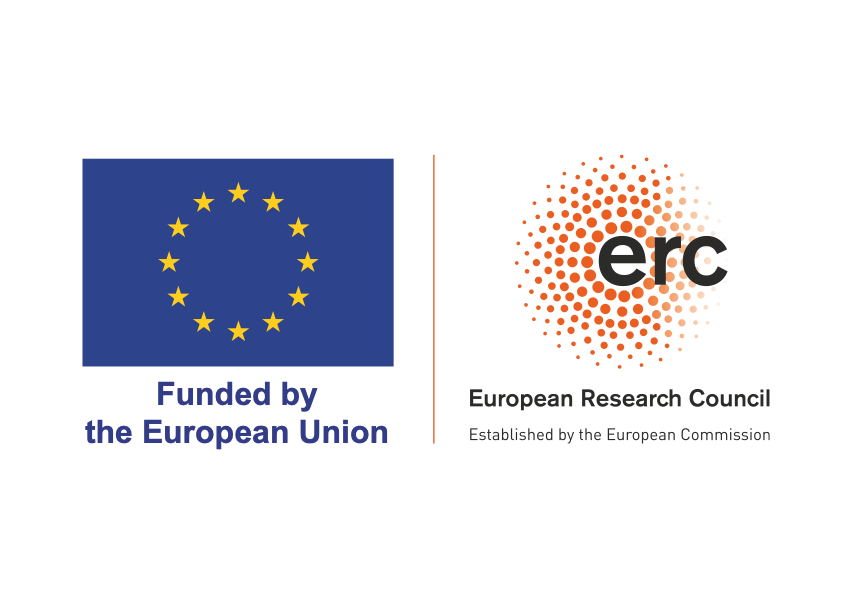}
\end{figure}


\appendix
\crefalias{section}{appendix}

\section{Finite element implementation of forward simulations}
\label{sec:FE_simulations}
In the absence of traction and body forces, the weak form of the forward simulations to generate the database reads
\begin{equation}\label{eq:weak_form}
 \int_{\mathcal{B}_0} \left[\left[\partial_{\bfF} \bar{W} - p J \bfF^{-\mathrm{T}} \right]: \nabla_0 \delta\bfu\, - \, \left[J-1\right]\delta p \right]\, \text{d}V,
= 0,
\end{equation}
where $\delta \bfu$ and $\delta p$ are test functions, and ${\mathcal{B}_0}$ is the material configuration.

The open-source finite element platform FEniCSx version \texttt{0.9.0} \citep{Baratta2023} is used to solve the weak form of the problem with a continuous Galerkin discretization of trial and test functions consisting of quadratic Lagrange polynomial basis functions for the displacement field and linear Lagrange polynomial basis functions for the pressure field (Taylor-Hood element). We use the library \texttt{ufl.diff} to compute $\partial_{\bfF} \bar{W}$ via symbolic differentiation. The finite element mesh has \qty{1016}{} nodes, \qty{2857}{} tetrahedra elements, and it mimics half the thickness of the sample and the displacement is fixed to zero on the middle-thickness symmetry plane.

\section{Experimental fingerprint measurement}
\label{sec:experimental_fingerprint_measurement}
As described previously, the fingerprints contain both reaction forces and displacements across multiple load steps. The raw experimental measurements must therefore be preprocessed to construct the experimental fingerprint. Specifically, the displacement fields obtained via DIC are downsampled to a predefined set of identification nodes located on an ellipse surrounding the initial notch (\cref{fig:geometry,fig:experiment}). The \texttt{Resample With Dataset} filter in ParaView is employed for this purpose, using the raw DIC measurements (coordinates and displacement values) and the coordinates of the ellipse points as inputs. In this work, $n_u = 50$ identification points were defined along the ellipse. The resampling is performed independently for each considered load step (clamp loading stage), displacement component (horizontal and vertical), and experimental repetition. Subsequently, the geometric mean across four experimental repetitions is computed at each identification point and load step. The resulting horizontal and vertical displacement components are then flattened across load steps and identification points to form the experimental displacement fingerprint vector $\bff_u$.

The construction of the experimental force fingerprint vector $\bff_R$ requires less preprocessing. Using the mapping between load-step indices and prescribed clamp displacements, the raw experimental force-displacement curves are evaluated at the desired load steps. When no data point is available exactly at a prescribed clamp displacement, the corresponding force value is obtained by linear interpolation along the experimental force-displacement curve. Analogous to the displacements, the (interpolated) forces in the horizontal and vertical directions are flattened across load steps to form $\bff_R$.

\bibliographystyle{elsarticle-harv}
\bibliography{bib_Moritz,bib_Miguel}

\end{document}